\documentclass[aps,amsmath,amssymb,prd,floatfix,preprint,superscriptaddress,nofootinbib,12pt]{JHEP3}
\usepackage{epsfig}
\usepackage{amsmath}
\usepackage{amssymb,amsfonts}
\usepackage{comment}
\usepackage{graphicx}
\usepackage{latexsym}
\usepackage{slashed}
\usepackage{xcolor}
\usepackage{subfigure}
\usepackage{subfig}
\usepackage[normalem]{ulem}
\let\normalcolor\relax
\def\be{\begin{equation}}
\def\ee{\end{equation}}
\def\bea{\begin{eqnarray}}

\def\eea{\end{eqnarray}}
\def\nn{\nonumber}

\def\half{\dfrac{1}{2}}

\def\6{\partial}

\def\normord{ {\scriptstyle {{\bullet}\atop{\bullet}}} }

 \def\nn{\nonumber\\[2pt]}       \def\be{\begin{eqnarray}}    \def\ee{\end{eqnarray}}
 \def\bi#1{\begin{itemize}\item[#1]}     \def\ei{\end{itemize}}  
   \def\^#1{\hat{#1}}

 \def\ffract#1#2{\raise .2 em\hbox{$\scriptstyle#1$}\kern-.3em/
                 \kern-.2em\lower .15 em \hbox{$\scriptstyle#2$}}
 
 \def\half{\frac{1}{2}}

\def\bmatrix{\begin{matrix}} \def\ematrix{\end{matrix}} \def\bpmatrix{\begin{pmatrix}}\def\epmatrix{\end{pmatrix}}
\def\bcenter{\begin{center}} \def\ecenter{\end{center}}

\def\lowerheightfig#1#2#3{\(\raise-#1\hbox{\includegraphics[height=#2]{#3}}\)}
\def\lowerwidthfig#1#2#3{\(\raise-#1\hbox{\includegraphics[width=#2]{#3}}\)}

\def\nn{\nonumber}

\def\hri#1#2{\href{http://arxiv.org/abs/#1}{[ArXiv:#1]#2}}

\def\hrj#1#2{\href{www.doi.org/#1}{#2}}

\title{Black Hole S-matrix for a scalar field}
\author{Panos Betzios$^{\star}$, Nava Gaddam$^{\dagger}$ and Olga Papadoulaki$^{\#}$\\
~\\
$^{\star}$ \href{http://www.apc.univ-paris7.fr}{APC, AstroParticule et Cosmologie}, Universit\'e Paris Diderot, CNRS/IN2P3, CEA/IRFU,
Observatoire de Paris, Sorbonne Paris Cit\'e,\\
 10, rue Alice Domon et L\'eonie Duquet, 75205 Paris
Cedex 13, France
~\\
$^\dagger$ \href{https://www.uu.nl/en/research/institute-for-theoretical-physics}{Institute for Theoretical Physics} \emph{and} Center for Extreme Matter and Emergent Phenomena, Utrecht University, 3508 TD Utrecht, The Netherlands \\
$^{\#}$ \href{https://www.ictp.it/}{International Centre for Theoretical Physics} \\
Strada Costiera 11, Trieste 34151 Italy.
}

\abstract{We describe a unitary scattering process, as observed from spatial infinity, of massless scalar particles on an asymptotically flat Schwarzschild black hole background. In order to do so, we split the problem in two different regimes governing the dynamics of the scattering process. The first describes the evolution of the modes in the region away from the horizon and can be analysed in terms of the effective Regge-Wheeler potential.  In the near horizon region, where the Regge-Wheeler potential becomes insignificant, the WKB geometric optics approximation of Hawking's is replaced by the near-horizon gravitational scattering matrix that captures non-perturbative soft graviton exchanges near the horizon. We perform an appropriate matching for the scattering solutions of these two dynamical problems and compute the resulting Bogoliubov relations, that combines both dynamics. This allows us to formulate an S-matrix for the scattering process that is manifestly unitary. We discuss the analogue of the (quasi)-normal modes in this setup and the emergence of gravitational echoes that follow an original burst of radiation as the excited black hole relaxes to equilibrium.}


\begin{document}

\section{Introduction}

Most recent approaches \cite{Strominger:1996sh, Mathur:2005zp, Almheiri:2019hni, Penington:2019npb} to resolve the information paradox either introduce unconventional\footnote{From the point of view of the low energy effective field theory.} degrees of freedom (such as strings) or employ the holographic duality that relates the bulk spacetime description with a non-gravitational strongly coupled quantum field theory that nevertheless `lives' in lower dimensions. There are of course various arguments and calculations in support of these approaches. However, certain physical questions such as the following, remain elusive:
\begin{itemize}
\item How exactly, and in which regime does the bulk effective field theory (EFT) fail?
\item Are such unconventional degrees of freedom strictly necessary to resolve the information paradox?
\item Since such degrees of freedom predominantly affect ultraviolet physics, and the asymptotic observer may choose to perform low energy experiments exclusively, why are they relevant in the first place?
\end{itemize}

An alternative approach is to rely on conventional degrees of freedom, but to incorporate a certain non-perturbative effect, that of strong gravitational interactions near the horizon. In the past, these have been seen to arise from gravitational backreaction of Dray and 't Hooft \cite{Dray:1984ha}. In a first quantised formalism, this effect leads to a departure from Hawking's calculation \cite{Hawking:1974sw, Hawking:1976ra}, due to a quantum mechanical scattering algebra realised by in and out going wavefunctions near the horizon \cite{Hooft:2015jea, Hooft:2016itl, Hooft:2016cpw, Hooft:2019nmf, Betzios:2016yaq, Betzios:2020wcv}.

As argued in \cite{Gaddam:2020mwe, Gaddam:2020rxb}, this departure is more generally attributed to soft graviton exchanges between the matter fields near the horizon\footnote{The interactions that lead to the non-perturbative amplitude have a coupling constant $M_{Pl}/M_{BH}$.
Nevertheless, it is worth noting that their resummation leads to a universal result, so that the near horizon scattering amplitude does not depend on the black hole mass, in Kruskal coordinates. The S-matrix described in the exponentiated Eddington-Finkelstein coordinates does gain dependence on the the mass of the black hole via the change in coordinates.}. This approach has therefore been confined to near horizon dynamics so far. Further away from the horizon, there is an effective gravitational potential due to the spacetime curvature that is ignored. Such soft graviton exchanges are familiar from perturbations around flat space, where collision energies are trans-Planckian \cite{Amati:1987uf, tHooft:1987vrq, Kabat:1992tb}. It is the purpose of this paper to combine the effects of this potential, together with the near horizon physics, and obtain a complete scattering description for the asymptotic observer far away from the horizon.

Spherical symmetry of the Schwarzschild horizon allows for the fluctuation equations for probe fields to be expressed in terms of a Schr\"{o}dinger-like equation, in a partial wave basis, with a radial potential named after Regge and Wheeler \cite{Regge:1957td}. This equation admits implicit solutions expressed as Heun functions, whose analytic properties are difficult to study~\cite{DLMF}. However, various approximations of the Regge-Wheeler potential have been employed to extract physics in different regimes of validity \cite{Iyer:1986np, Iyer:1986nq, Motl:2003cd}; see also the references \cite{Kokkotas:1999bd, Konoplya:2011qq}. In particular, focusing on the region near the maximum of the potential, it was shown in \cite{Iyer:1986np, Iyer:1986nq} that approximate solutions (which nevertheless resemble the true solutions rather closely) to the Regge-Wheeler equation can be found. Using these, quasi-normal modes of the Schwarzschild black hole have been analytically calculated, to find an agreement of within $0.7\%$ of the numerically calculated values. Most studies of these quasi-normal modes assume a fixed background geometry and any potential consequences of black hole evaporation are ignored, by imposing purely in-going boundary conditions at the horizon. A scattering relation between the near horizon ingoing and outgoing modes will of course modify this result. In combining the effects of the Regge-Wheeler potential with this near horizon scattering relation, in this paper, we will be able to explicitly calculate the S-matrix for the asymptotic observer as well as the modified resonances that describe the analogue of quasi-normal modes. We should also emphasise that the present model has two natural timescales, the ``scrambling time'' of the near horizon scattering of the order $2 G M \log G M^2$ and a longer timescale that governs the time it takes for modes to eventually escape to asymptotic infinity. When the near horizon dynamics is modified and one can no longer impose purely infalling boundary conditions at the horizon, gravitational `echoes' appear and show up as repetitive peaks arising from the backscattering between the near horizon and the Regge-Wheeler potential~\cite{Cardoso:2016rao, Cardoso:2016oxy, Cardoso:2019rvt, Abedi:2016hgu, Wang:2018gin}\footnote{There is, however, a fundamental difference compared to these works considering effectively classically horizonless geometries. In the present article there is a black hole horizon, but equipped with modified boundary conditions for the modes of the external observer; these boundary conditions take both quantum mechanics and gravitational backreaction into account.}. This echo time can be calculated, and interestingly, it is indeed related to the scrambling time \cite{Abedi:2016hgu, Wang:2018gin}. Our model has no freely adjustable parameters and hence can potentially serve as a good template model in the search of such gravitational echoes.

In this article, we present a construction of the complete scattering matrix that incorporates both the effects alluded to above, that are associated to the near horizon gravitational interactions and to the Regge-Wheeler gravitational potential. This is done by imposing appropriate matching conditions for the modes (associated to the \emph{transfer} T-matrix) entering the Regge-Wheeler potential and those arising in the near horizon gravitational dynamics (associated to the near horizon \emph{scattering} matrix). Such \emph{connection} problems in the context of black holes have been studied in the past, a selection of these studies can be found in \cite{Betzios:2017dol, Betzios:2018kwn, Porfyriadis:2018yag, Bena:2019azk, Bena:2020yii}. We perform this matching in two cases:

\paragraph{Two-sided black hole} We first analyse the problem on the extended Kruskal manifold. Classically, there is no propagation for modes from one exterior to the other, as the inserted shockwaves arise from a stress tensor $T_{VV} = p_{in} \delta^{(3)}\left(U, \Omega\right)$ that satisfies the null energy condition. After quantisation, however, probability for such propagation arises; this process requires quantum mechanics and gravity to simultaneously operate and is mediated by a non-perturbative exchange of soft gravitons. Furthermore, the resulting S-matrix is unitary on the extended Kruskal manifold. The matching conditions, therefore, result in a matrix equation capturing both exterior regions. The resulting matrix that combines the dynamics near the horizon and the Regge-Wheeler potential, is manifestly unitary. 

\paragraph{Single-sided black hole} The two-sided case can be reduced to a single-sided case via an antipodal identification advocated by 't Hooft \cite{Hooft:2016itl}, or via its near horizon quantum analog proposed by the present authors \cite{Betzios:2020wcv}. A natural question is whether a single-sided solution so obtained, is representative of a black hole formed in a realistic collapse. For the purposes of questions of interest in this article, we argue that it is. Our reasoning is as follows:

In a realistic problem, after evaporation, all modes received at $\mathcal{I}^+$ appear to emanate from the near horizon region. For late-time modes, this is predominant and the details of collapse appear hidden in the near horizon region. Hawking's geometric optics approximation would require us to continue these modes past the collapsing star and into asymptotic past infinity $\mathcal{I}^-$. Instead, observing that the large blueshift and the relative boost between the in and outgoing modes cause strong gravitational interactions near the horizon, these necessarily affect the continuation to past inifnity; information from ingoing modes is passed on to the outgoing ones via these interactions. In fact these interactions expressly invalidate the geometric optics approximation that assumes no interaction. The unitary map so obtained from the gravitational backreaction effect yields a Bogoliubov transformation between the in and outgoing modes in the near horizon region. The propagation of these near horizon modes to null infinity is necessarily obstructed by the Regge-Wheeler potential; this motivates and necessitates our matching conditions in the present article. What follows is a unitary scattering matrix (using standard formulae which we document in Appendix \ref{Bogoliubov}). Therefore, we reason that if focus is on the late time scattering matrix of a scalar field propagating on a large Schwarzschild background\footnote{By `late times', we mean long after the size of the black hole has outgrown Planck size. The black hole returns to Planck size in the very last moments of evaporation. Such final stages of evaporation are also excluded in our work.}, we may legitimately neglect any effects coming from the process of the initial collapse of the star that formed the black hole in the first place, provided the near horizon gravitational interactions are appropriately accounted for; these restore unitarity in the effective field theory regime\footnote{Unitarity for any ultraviolet or string theory modes must also be shown by extending the present formalism, as in \cite{Amati:1987wq, deVega:1988ts, Lousto:1996zk}.} we work in. 

Even though we do not have access to true microscopic Planckian dynamics, the quantum analog of the antipodal identification yields a discrete chaotic spectrum for states at a Planckian distance from the horizon, as a boundary condition similar to a ``cavity resonator" model \cite{GutzwillerBook, Gutzwiller, Lax, Faddev, Savvidy:2018ffh, Betzios:2020wcv}. The coupling and coexistence of a discrete spectrum with the continuous spectrum of scattering states, from the farther region near the Regge-Wheeler potential, is generically expected to result in complicated features such as the so-called Fano resonances \cite{Fano:1961zz}. Nevertheless, if the spectral gap between consecutive levels describing the near horizon microstates is extremely small (as was found to be the case in \cite{Betzios:2020wcv}), we may simply neglect such effects and merely treat the discrete spectrum as an approximate continuum. This is what we do in the present article. Our treatment of the quantum black hole, therefore, resembles a very complicated nucleus or atom. While there is an internal discrete densely spaced chaotic spectrum, we may consider the scattering states of a scalar field on this ``black hole nucleus" and treat them in a continuous fashion.\footnote{A more detailed description may be provided by the scattering of localised wavepackets which we briefly analyse in appendix~\ref{Wavepackets}.} In this paper, we argue that quantum chaotic properties are encoded in the Bogoliubov coefficients that relate the ingoing and outgoing modes of the second quantised theory.

The structure of this paper is the following. In Section \ref{reviewSmatrix}, we review the scattering matrix approach to black holes. Thereafter, we review the Regge-Wheeler equation, its solutions, and introduce the matching conditions in section~\ref{matching}. These matching conditions are applied to the two-sided and one-sided black holes in the sections that follow. In section~\ref{singleexteriorSmatrix}, we also discuss the corrections to the quasi-normal modes (QNM's) that arise from our matching conditions and in the following section \ref{ECHOES} we interpret them as gravitational echoes following the main burst of radiation as the black hole relaxes to equilibrium. We end with the conclusions and a discussion on the problems for the future, in section \ref{Conclusions}.

\section{A short review}\label{reviewSmatrix}

To set the stage for an analysis of the S-matrix, we first review Hawking's original approach \cite{Hawking:1974sw,Hawking:1976ra} which suggested that black holes radiate in a thermal fashion. It concerns a black hole with a single exterior formed by collapse. The main physical idea in these works is that when the modes on future null infinity $\mathcal{I}^+$ are traced into the past towards the horizon, they suffer a blueshift and pile up in the near horizon region. This validates the use of a WKB geometric optics approximation in determining the evolution of the various modes on a black hole background. The WKB approximation employed, and the resulting Bogoliubov transformations that relate the modes on past null infinity $\mathcal{I}^-$  to those on $\mathcal{I}^+$, hold only long after the black hole has formed. Therefore, such an approach cannot model the complete collapse and evaporation process.
 
Consider a massless scalar field that admits the following mode expansion
\be
\hat{\Phi} \, = \, \sum_{l m} \int_0^\infty \frac{d \omega}{\sqrt{2 \pi \omega}} \left( \hat{a}_{\omega l m} \Phi_{\omega l m} \, + \,  \hat{a}_{\omega l m}^{\dagger } \Phi_{\omega l m}^{*}  \right)\, ,
\ee
in terms of solutions (modes) $\Phi_{\omega l m}$ to the scalar equation of motion $\Box_g \Phi = 0 $ on the black hole background. The operators $\hat{a}_{\omega l m}$ and $\hat{a}_{\omega l m}^{\dagger }$ are creation and annihilation operators corresponding to each mode, respectively. The ingoing modes at $\mathcal{I}^-$ depend on the affine parameter $v$, while the outgoing modes at $\mathcal{I}^+$ on the parameter $u$.

The WKB approximation leads to the following\footnote{It is customary to use a rescaled field $\phi$ at $\mathcal{I}^\pm$  (the so called \emph{image} of $\Phi$), that is given by $\Phi = \phi/r$ as $r \rightarrow \infty$. The rescaled field $\phi$ is the one with an appropriate Klein-Gordon inner product at null infinity and determines the in/out data.} relation between the asymptotic solutions at $\mathcal{I}^-$ and $\mathcal{I}^+$
\be\label{WKBrel}
\phi_{in}(v, \Omega) = \phi_{out}(u(v), \Omega^P) \, , \qquad u(v) = v_0 - 4 G M \log \left(\frac{v_0 - v}{4 G M} \right) \, , \quad v < v_0 \, .
\ee
In this expression $\Omega^P$ is the antipodal point of $\Omega$ on the two-sphere, and $v_0$ is a constant parametrising the location of the mode on $\mathcal{I}^-$ that is to coincide with the future horizon. This WKB coordinate relation gives rise to a Bogoliubov map between ingoing and outgoing modes of the form
\bea
\phi^{out}_{\omega l m} = \int_0^\infty d \omega' \left(A_{\omega \omega'} \, \phi^{in}_{\omega' l m} \, + \, B_{\omega \omega'} \, \phi^{* \, in}_{\omega' l m}  \right) \, ,
\eea
where $\phi^{in}_{\omega' l m}$ are the modes with support on $\mathcal{I}^-$ and $\phi^{out}_{\omega l m}$ are modes with support on $\mathcal{I}^+$. In this specific approximation, the Bogoliubov coefficients are found (up to an irrelevant overall phase) to be
\begin{align}
A_{\omega \omega'} ~ &= ~ e^{- i(\omega' - \omega) v_0} \frac{e^{2 \pi G M \omega} \Gamma(1 - i 4 G M \omega)}{2 \pi \sqrt{\omega(\omega' + i \epsilon)}} \nonumber \\
B_{\omega \omega'} ~ &= ~ e^{ i(\omega' + \omega) v_0} \frac{e^{- 2 \pi G M \omega} \Gamma(1 - i 4 G M \omega)}{2 \pi \sqrt{\omega(\omega' - i \epsilon)}} \, .
\end{align}
These relations give rise to a thermal density matrix describing the outgoing quanta. In particular assuming an ingoing pure state (for example the vacuum), we find that the $out$-quanta will be thermally distributed\footnote{There are corrections to this result arising from the so-called greybody factors, due to the effective Regge-Wheeler potential that we analyse in section \ref{ReggeWheeler}, but these do not invalidate the thermal nature of the outgoing radiation in Hawking's computation.} in terms of a Bose-Einstein distribution.
One shortcoming of this approach was already pointed out by Hawking himself \cite{Hawking:1974sw,Hawking:1976ra} (see also \cite{Kiem:1995iy}). It is that this calculation does not take into account the gravitational backreaction of ingoing modes on the outgoing ones and on the geometry itself. This is a substantial effect from the point of view of the late time asymptotic external observer due to the blueshift of the modes as their evolution is traced backwards in time towards the horizon. This then results in a substantial lightcone stress energy tensor that affects any scattering process from the point of view of such an observer that is accelerating away from the horizon. An equivalent way of seeing this is by computing the near horizon stress energy tensor in the vacuum state perceived by the observer at $\mathcal{I}^+$. This vacuum state is called the Boulware vacuum \cite{Boulware:1974dm} and the resulting expression for the stress energy tensor is found to diverge\footnote{A similar divergence is found in the past horizon of the Unruh vacuum \cite{Unruh:1976db} as discussed in appendix \ref{Vacua}.} precisely due to this blueshift effect. 

Gravitational backreaction can be computed classically by solving the nonlinear Einstein's equations sourced by stress tensors corresponding to light-like shockwaves \cite{Aichelburg:1970dh, Dray:1984ha, Sfetsos:1994xa}. Such non-linearities are known to arise from an eikonal resummation of quantum gravitational interactions \cite{Amati:1987uf, tHooft:1987vrq, Kabat:1992tb}. In a first quantised formalism, these interactions have been studied over the years \cite{Hooft:2015jea, Betzios:2016yaq, Hooft:2016itl, Betzios:2020wcv}, and we summarise the main results of these works in Appendix \ref{FirstQuant}. More recently, it was shown that there is a `black hole eikonal phase' of quantum gravity where gravitational interactions near the horizon can be seen to arise from soft graviton exchange processes between in and out going modes near the horizon \cite{Gaddam:2020mwe, Gaddam:2020rxb}. This is in a quasi-adiabatic approximation where the black hole is much larger than Planck size and is slowly evaporating. In particular, in this approximation, the rate of change of the black hole (mass, for instance) is slower than the timescale associated to the scattering process of the individual modes in the near horizon region; the latter is sometimes called scrambling time. With these assumptions, in a partial wave basis, the near-horizon scattering matrix can be calculated to be \cite{Gaddam:2020mwe, Gaddam:2020rxb}
\begin{equation}\label{eqn:amplitude}
\langle P_{out}; \ell m | \hat{S}_{horizon} | P_{in}; \ell m \rangle ~ = ~ 4 P_{in} P_{out} \exp \left( i \frac{8 \pi G}{l^{2} + l + 2} P_{in} P_{out}\right) \, ,
\end{equation}
where $P_{in/out}$ are lightcone momenta of ingoing and outgoing particles on the $S^2$, and $l, ~ m$ denote the spherical harmonics. This amplitude is intriguing since it arises in a resummation of a perturbative expansion of ladder diagrams in the coupling constant $G/R^{2}$ and therefore trivialises when either gravity or quantum mechanics is switched off. The amplitude is also consistent with the one obtained via first quantisation methods \cite{Hooft:2015jea, Betzios:2016yaq, Hooft:2016itl, Betzios:2020wcv}\footnote{The second quantised amplitude \eqref{eqn:amplitude} contains off-shell graviton fluctuations that result in a discrepancy of the prefactor inside the exponential, compared to the first quantised results.}. It is of course possible that additional string theoretic corrections could become important for smaller black holes. Regardless of potential corrections to the amplitude \eqref{eqn:amplitude} in a UV complete theory, it is evidently unitary already at the level of the effective field theory.

In contrast to Hawking's WKB result \eqref{WKBrel} that gave rise to a Bogoliubov map for asymptotic states,  the amplitude \eqref{eqn:amplitude} provides for a \emph{near horizon mode relation} that can be used to describe the scattering of a massless scalar field. As we review in Appendix \ref{Bogoliubov}, given such a scattering amplitude for the modes, the Bogoliubov coefficients relating the near horizon in and outgoing modes can be extracted. We will show in this paper that this mode relation modifies Hawking's WKB approximation and gives rise to different Bogoliubov coefficients that result in unitary amplitudes.

All the dynamics of the scattering process take place in the external region of the black hole and relies on the presence of a horizon provided collision energies satisfy $M_{BH} \gg E \gg M^{2}_{P}/M_{BH}$. These energies are very low for large semi-classical black holes, ensuring that there is no trans-Planckian problem and that infalling observers do not encounter firewalls \cite{Gaddam:2020mwe, Gaddam:2020rxb}\footnote{By this we mean that the window of validity of the computation is very large and contains low energy observers. Of course, a very energetic pulse with $E \sim O(M_{BH})$ would greatly disturb the local geometry near the horizon.}. See also \cite{Pasterski:2020xvn} for a similar perspective.

The central goal then, is to show how the near horizon modes can be traced all the way up to asymptotic infinity to obtain a complete S-matrix including the effects of the background geometry coming from the gravitational Regge-Wheeler potential. The scattering matrix thus obtained is a unitary matrix.

\section{Regge - Wheeler equation}\label{ReggeWheeler}

Both Hawking's calculation and the subsequent improvements in \cite{Hooft:2015jea, Betzios:2016yaq, Gaddam:2020mwe, Gaddam:2020rxb} that include gravitational interactions need to account for the presence of an effective gravitational potential for the propagating modes on the Schwarzschild background. This potential gives rise to additional reflection and transmission coefficients for the scattering problem; these are related to the greybody factors of the black hole. The Klein-Gordon equation for a massless field on the Schwarzschild background is
\be\label{RW1}
\Box_g \Phi ~ = ~ 0 \, ,
\ee
where $g$ refers to the Schwarzschild metric. Using the ansatz $\Phi = r^{-1}\phi_{l m}(r, t) Y_{l m}(\Omega)$ for the scalar field modes, the above equation simplifies to
\be\label{RW2}
\left(\frac{\partial^2}{\partial r_*^2} - \frac{\partial^2}{\partial_t^2} - V_l(r) \right) \phi_{l m}(r,t) = 0 \, .
\ee
Defining the Regge Wheeler coordinate as 
\be\label{RW3}
r_* ~ = ~ r + 2 G M \log(r/2GM - 1) 
\ee
pushes the location of the horizon from $r = 2 GM$ to $r_* = - \infty$. Expanding the field in frequency Fourier modes $\phi_{l m}(r,t) = e^{-i\omega t} \phi_{l m}(r, \omega)$, we find
\be\label{RW4}
\left(\frac{\partial^2}{\partial r_*^2}  + \omega^2 - V_l(r) \right) \phi_{l m}(r, \omega) = 0 \, ,
\ee
where the effective potential is then\footnote{Modes of different spin can be treated in a similar fashion, see \cite{Chandrasekhar:1985kt} and references therein.}
\be\label{RW5}
V_l(r) ~ = ~ \left( 1 - \frac{2 GM}{r} \right) \left[\frac{l(l+1)}{r^2} - \frac{2GM}{r^3} \right] \, .
\ee
The maximum of this potential is at $r \approx 3 GM$ (the light-ring)~\cite{Cardoso:2008bp}.

The physics of the potential is straightforward: a part of a wave emanating from $\mathcal{I}^-$ is transmitted across the potential towards the horizon, and a part is reflected back. Similar fate awaits modes emanating from the horizon that try to escape to $\mathcal{I}^+$; some succeed and some are reflected back. Finally, quasinormal modes (QNM's) are classically interpreted as marginally trapped waves at this light-ring that are leaking out of it~\cite{Cardoso:2008bp}. Unfortunately, analytic control over solutions to the Regge-Wheeler equation is intractable. However, various approximations have been employed to study the resulting physics; see \cite{Iyer:1986np, Iyer:1986nq, Motl:2003cd} and the reviews \cite{Kokkotas:1999bd,Konoplya:2011qq}.

\begin{figure}[t]
\centering
\includegraphics[width=0.7\textwidth]{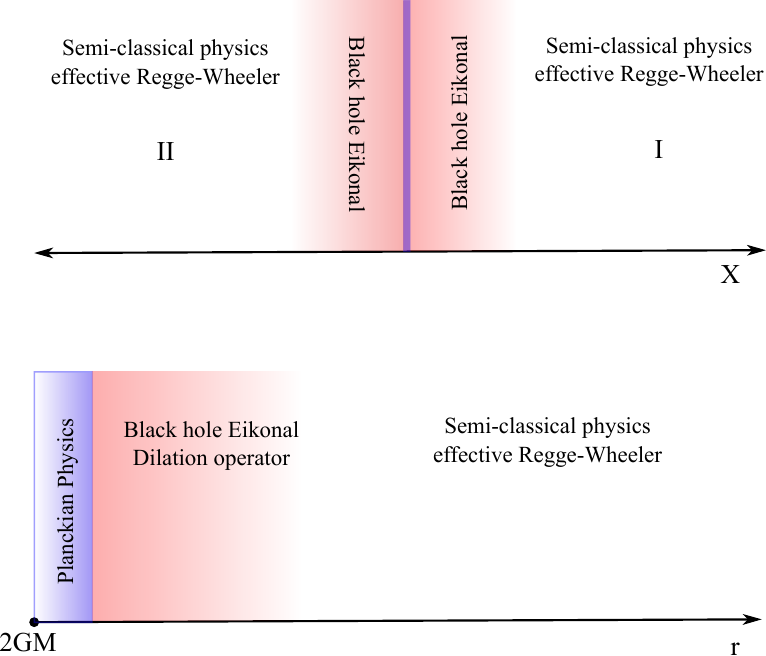}
\caption{Propagating fields far from the horizon evolve semi-classically in a fixed background and the physics is largely governed by the effective Regge-Wheeler potential. In the near horizon region $\ell_{P} \ll r - 2GM \ll GM$, the Regge-Wheeler potential vanishes. The dominant effect is that of gravitational interactions governed by the black hole eikonal phase. The in and out modes are related via the scattering kernel \eqref{eqn:amplitude} and they must be matched with the modes near the Regge-Wheeler potential for continuity. An antipodal identification in the effective theory was proposed by 't Hooft to go from the two-sided black hole to the single-sided one. It is also possible to define a quantum generalisation of it that is consistent with the commutation relations \eqref{B8}. This has the added advantage that the spectrum of microstates is now chaotic. Since the spectral spacing is extremely dense, we shall not consider the effects of such a discretisation in the present article.}
\label{fig:gluing}
\end{figure}

Since the strong gravitational interactions dominate the near horizon physics where the Regge-Wheeler potential vanishes, both effects can be disentangled and simultaneously studied. What this entails is a matching of the near horizon modes of the black hole eikonal scattering with the scattering solutions of the Regge-Wheeler equation which affect the physics further away from the horizon. This idea of matching the solutions is pictured in fig. \ref{fig:gluing}, both for the two-sided and one-sided black hole. Such a disentangled study of the two effects and a combination of them via a mere matching condition is only valid for large (compared to Planck size) semi-classical black holes where there is a large separation of scales. There are three length scales of interest that are well separated:
\begin{itemize}
\item The first is the regime a Schwarzschild radius away from the horizon where $r-2GM/c^2 \sim \mathcal{O}(GM/c^2)$. Here, semi-classical effects of quantum fields on a fixed background dominate the physics. We may reliably study the effect of the fixed Regge-Wheeler potential on the quantum fluctuations. 
\item The next is the near horizon region where $\ell_{P} \ll r-2GM/c^2 \ll O(GM/c^2)$ where strong gravitational effects gain importance. This is the region where the information paradox arises from, and one may have feared that effective field theory may prove too limited or non-renormalisability of gravity too challenging to address this regime. As was recently shown in \cite{Gaddam:2020mwe, Gaddam:2020rxb}, effective field theory indeed captures unitary physics in this regime provided a certain non-perturbative resummation (in $M_{P}/M_{BH}$) of soft graviton exchanges is taken into account. Non-renormalisability does not pose a threat owing to the emergent black hole eikonal phase.
\item Finally, there is the Planckian regime $r-2GM/c^2 \sim O(\ell_{P})$ where Planckian physics and large momentum transfer effects dominate. However, it is plausible that this physics can effectively be described in the low energy theory using the non-commuting scattering algebra of \cite{Hooft:2015jea, Betzios:2016yaq} via an effective boundary condition \cite{Hooft:2016itl, Betzios:2020wcv}; this effective boundary condition results in a rich chaotic spectrum of black hole microstates (as probed by an external observer), and an important open problem is if these account accurately for the entropy of the black hole. The microscopic theory from which such a boundary condition may descend is unknown.\footnote{Some preliminary efforts to construct such a microscopic chaotic model can be found in \cite{Banks:2015mya, Betzios:2016yaq, Banks:2020zrt}}
\end{itemize}

Extending this approach to fields with additional degrees of freedom such as charge or spin may be incorporated into the gravitational interactions near the horizon \cite{Gaddam:2020mwe, Gaddam:2020rxb} before matching them with the solutions of the appropriate Regge-Wheeler or Zerilli equation for a spin-$s$ field \cite{Chandrasekhar:1985kt, Martel:2005ir}. 

\subsection{Matching modes in the near horizon region}\label{matching}
For what is to follow, it is convenient to use the ingoing/outgoing Eddington-Finkelstein coordinates $u= t - r_*, \,  v = t+ r_*$ to describe the evolution of the modes. The ingoing positive frequency wavepackets describing solutions to the Klein-Gordon equation~\ref{RW1}, expressed in ingoing Eddington-Finkelstein coordinates on $\mathcal{I}^-$ are given by
\be
\Phi^{in}(v,r,\Omega) ~ = ~ \frac{\phi^{in}(v,r,\Omega)}{r}  \sim  Z^{in \, (as)}_{\omega l m} \frac{e^{- i \omega v}}{r} Y_{l m }(\Omega) \, ,
\ee
whereas the outgoing ones on $\mathcal{I}^+$ are given by
\be
\Phi^{out}(u,r,\Omega) ~ = ~ \frac{\phi^{out}(u,r,\Omega)}{r}  \sim  Z^{out \, (as)}_{\omega l m} \frac{e^{- i \omega u}}{r} Y_{l m }(\Omega) \, .
\ee
In order to completely specify the evolution, we also need the near horizon expansion of the modes that takes a similar form. The solution of the Regge-Wheeler equation then provides a connection (transfer) matrix for the coefficients between solutions in the near horizon and the asymptotic region at infinity
\begin{align}
\phi^{(hor)} ~ &\sim ~ Z^{out \, (hor)}  \, e^{i \omega r_*}  \, + \, Z^{in \, (hor)} \, e^{-i \omega r_*}  \, ; \quad r_* \rightarrow - \infty \, , \nn \\
\phi^{(as)} ~ &\sim ~ Z^{out \, (as)} \, e^{i \omega r_*} ~~ + \, Z^{in \, (as)}\,  e^{-i \omega r_*} \, ; ~ \quad r_* \rightarrow + \infty
\, .
\end{align}
These general asymptotic expansions of course require boundary conditions that depend on the physical problem under study. For example, a boundary condition commonly employed is that of purely infalling modes at the horizon
\begin{align}\label{quasi}
\phi^{(hor)} ~ &\sim ~ e^{-i \omega r_*} && r_* \rightarrow - \infty \, , \nn \\
\phi^{(as)} ~ & ~ \sim Z^{out \, (as)} e^{i \omega r_*} + Z^{in \, (as)} e^{-i \omega r_*} && r_* \rightarrow + \infty \, .
\end{align}
The coefficients in this case are found to obey $1+ |Z^{out \, (as)}|^2 = |Z^{in \, (as)}|^2$. From \eqref{quasi}, it is evident that a part of the wave is lost, since nothing returns from the horizon. 

On the other hand, we know that quantum mechanically the black hole radiates and this means that there are also modes that can exit from the near horizon region to infinity. In addition, as was emphasized in~\cite{Dray:1984ha,Hooft:2015jea, Hooft:2016itl, Hooft:2016cpw, Hooft:2019nmf, Betzios:2016yaq, Betzios:2020wcv,Gaddam:2020mwe, Gaddam:2020rxb} and reviewed in section~\ref{reviewSmatrix} of the present article, the ingoing modes interact with the outgoing ones resulting in a near horizon scattering relation between them. This means that it is possible and necessary to take into account both ingoing and outgoing modes that enter and exit the near horizon region. The next step then is to impose a matching condition between the Regge-Wheeler modes and these black hole eikonal modes in the near horizon region. The scattering matrix \eqref{eqn:amplitude} then provides for the following mode relation
\be\label{mode_rel}
Z^{out \, (hor)}_{\omega l m} \,  = \, \hat{S}_{horizon} \, Z^{in \, (hor)}_{\omega l m} \, .
\ee
In the case of a two-sided black hole, the Kruskal modes in \eqref{eqn:amplitude} must be expressed in terms of Eddington-Finkelstein modes on either exterior, which means that the above equation is a matrix equation as we will show in the next section \ref{twoexteriors} in greater detail. For now, we first note that the Klein-Gordon equation near the horizon is effectively reduced to one on Rindler space in lightcone coordinates. This implies that the solution to the near horizon wave equation splits in the following way
\begin{equation}
\phi^{(hor)}_{lm}\left(u,v\right) ~ = ~ \phi^{in (hor)}_{lm}\left(v\right) + \phi^{out (hor)}_{lm}\left(u\right) \, ,
\end{equation}
with the ingoing and outgoing modes given as
\be\label{inouthormodes}
\phi^{in}_{\omega l m}  ~ =  ~ Z^{in \, (hor)}_{\omega l m} e^{- i \omega v} \, , \qquad \phi^{out}_{\omega l m} ~  = ~ Z^{out \, (hor)}_{\omega l m} e^{-i \omega u} \, .
\ee
Written in Kruskal coordinates, these modes can be expressed as \footnote{An appropriate rescaling of the fields to ensure orthonormality with respect to these coordinates needs to be accounted for \cite{Betzios:2016yaq, Betzios:2020wcv}.}
\be
\Psi^{in}_{E l m} (V) = \frac{1}{\sqrt{2 \pi \hbar}} A^{in}_{E l m} \frac{1}{|V|^{\half - i E/ \hbar}} \, , \quad \Psi^{out}_{E l m} (U) =  \frac{1}{\sqrt{2 \pi \hbar}} A^{out}_{E l m} \frac{1}{|U|^{\half + i E/ \hbar}}  \, . 
\ee
where we reintroduced the physical units for energies $E/\hbar = 4 G M \omega$. These are the even modes under $V\rightarrow - V$ and $U \rightarrow - U$; there are corresponding odd modes with an additional sign. It is evident now that these modes are the eigenfunctions of the Dilation operator in the first quantised formalism studied in \cite{Betzios:2016yaq, Betzios:2020wcv} and briefly reviewed in Appendix \ref{FirstQuant}. The only complication is that the problem in Kruskal coordinates contains two near horizon exteriors. We tackle this by simply introducing two sets of near horizon Rindler modes in section \ref{twoexteriors}.

Therefore, we conclude that our matching condition is to be imposed on the solutions of two different effective Schroendinger problems: the Regge-Wheeler problem governing the region away from the horizon, and the Dilation Hamiltonian describing the near horizon evolution. This matching is made possible since the solutions of these two scattering problems take exactly the same form in a non-vanishing overlapping region near the black hole horizon. The near horizon scattering relations of the black hole eikonal phase can be written in terms of the eigenmodes of the Dilation operator; we are free to choose either formalism.

\begin{figure}[t]
\centering
\includegraphics[width=0.5\textwidth]{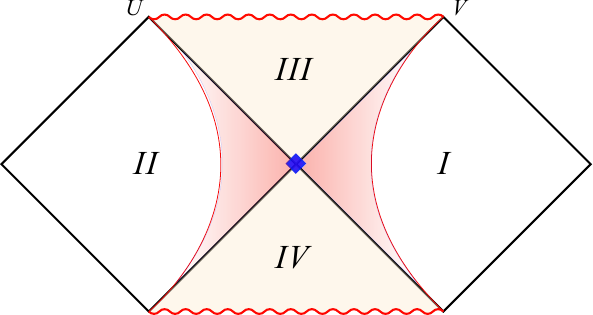}
\caption{The various regions, depicted on the extended Penrose diagram. In the white region the Regge-Wheeler potential dominates, while in the pink region the near-horizon gravitational interactions mediated by soft gravitons take over. The blue region is at a Planckian distance from the horizon, where the non-commutative structure of the backreaction algebra~\eqref{B8} would become important, resulting in a highly dense chaotic discrete spectrum of states eqn.~\ref{densespectrum}. In this work we do not analyse any such Planckian effects.}
\label{fig:Regions}
\end{figure}

\section{Asymptotic S-matrix for two-sided black holes}\label{twoexteriors}

As explained in Section \ref{reviewSmatrix}, the near horizon scattering kernel \eqref{eqn:amplitude} provides for a Bogoliubov relation between the near horizon ingoing and outgoing modes. Matching these modes with those transferring across the Regge-Wheeler potential gives us a Bogoliubov relation between the asymptotic ingoing and outgoing modes for the massless scalar field. In the case of a two-sided black hole, the scattering relation is in fact a $2 \times 2$ matrix \cite{Dray:1984ha, Hooft:2015jea, Betzios:2016yaq, Hooft:2016itl, Betzios:2020wcv}. Consequently, this results in a Bogoliubov matrix which can be found by forming appropriate inner products as shown in appendix \ref{Bogoliubov}. From now on, we introduce an extra subscript $\pm$ to keep track of the two exteriors.

Using the partial wave scattering kernel \eqref{eqn:amplitude} and the near horizon Eddington-Finkelstein modes \eqref{inouthormodes}, one finds the following matrix relation between them \cite{Hooft:2015jea, Betzios:2016yaq}
\begin{equation}\label{NearhorBogo}
\left( \begin{array}{c}
Z^{out \, (hor)}_{\omega l m + } \\
Z^{out \, (hor) }_{\omega l m - }  \end{array} \right) ~ = ~  \left( \begin{array}{cc}
S^{horizon}_{++}\left(\omega \right) & S^{horizon}_{+-}\left(\omega \right) \\
S^{horizon}_{-+}\left(\omega \right) & S^{horizon}_{--}\left(\omega \right) \end{array} \right) \left( \begin{array}{c}
Z^{in \, (hor)}_{\omega l m + } \\
Z^{in \, (hor)}_{\omega l m - } \end{array} \right) \, .
\end{equation}
The matrix is explicitly given by ($\lambda_l = 8 \pi G /R^2(l^2+l+2)$)
\begin{align}\label{eqn:SMatrix2b}
S^{horizon}\left(\omega \right) ~ = ~ \dfrac{1}{\sqrt{2\pi}} \Gamma\left(\dfrac{1}{2} - i \, 4 G M \omega \right) e^{-i 4 G M \omega  \log\lambda_l} \left( \begin{array}{cc}
e^{- i\frac{\pi}{4}} \, e^{- 2 \pi G M \omega }  &  e^{ i\frac{\pi}{4}} \, e^{  2 \pi G M \omega }  \\
e^{ i\frac{\pi}{4}} \, e^{  2 \pi G M \omega }  & e^{- i\frac{\pi}{4}} \, e^{- 2 \pi G M \omega }   \end{array} \right) \, .
\end{align}

On the other hand, the solutions to the Regge-Wheeler equation in the near horizon and asymptotic regions are related by two copies of the transfer matrix \cite{Iyer:1986np, Iyer:1986nq, Konoplya:2011qq}, one for each exterior
\begin{equation}\label{ReggeWheelerconnection}
\left( \begin{array}{c}
Z^{out \, (hor)}_{\omega l m \pm } \\
Z^{in \, (hor) }_{\omega l m \pm }  \end{array} \right) ~ = ~  \left( \begin{array}{cc}
T^{RW}_{11} & T^{RW}_{12} \\
T^{RW}_{21} & T^{RW}_{22} \end{array} \right) \left( \begin{array}{c}
Z^{out  \, (as)}_{\omega l m \pm } \\
Z^{in \,  \, (as)}_{\omega l m \pm } \end{array} \right) \, ,
\end{equation}
where
\be\label{eqn:TmatrixConditions}
T^{RW*}_{11} =T^{RW}_{22}\,,\quad  T^{RW*}_{12} =  T^{RW}_{21}\,,\quad \det{T^{RW}}= 1 \, .
\ee
These last relations above ensure the reality of the potential and show the fact that $T^{RW}$ is a transfer matrix. The matrix elements of the transfer matrix can also be expressed in terms of the reflection ($\mathcal{R}$) and transmission ($\mathcal{T}$) amplitudes
\begin{equation}\label{RWreflectiontransmission}
T^{RW} \, = \, \left( \begin{array}{cc}
T^{RW}_{11} & T^{RW}_{12} \\
T^{RW *}_{12} & T^{RW *}_{11} \end{array} \right) ~=~ \left( \begin{array}{cc}
\mathcal{T}^{-1} & \mathcal{R} \mathcal{T}^{-1}  \\
- \mathcal{T}^{-1} \mathcal{R} & (\mathcal{T}^*)^{-1} 
\end{array} \right) \, ,
\end{equation}
with $|\mathcal{R}|^2+|\mathcal{T}|^2 = 1$.
While exact analytic expressions for the elements of the transfer matrix are difficult to obtain, approximate ones are tractable. In particular, assuming that the turning points of the potential are close to each other (implying that the scattering energies are near the tip of the maximum of the potential), a WKB approximation results in the following matrix elements \cite{Iyer:1986np, Iyer:1986nq, Konoplya:2011qq}
\be\label{T1}
T^{RW} = \begin{pmatrix}
\frac{P^{-2} (2\pi)^{\frac{1}{2}}}{\Gamma (-\nu)} & -e^{i\pi\nu} \\
e^{i\pi\nu} & \frac{i P^{2} e^{i\pi\nu} (2\pi)^{\frac{1}{2}}}{\Gamma (\nu +1)}
\end{pmatrix}.
\ee
Here, the parameter $\nu(\omega)$ is given by
\be\label{T2}
i \left(\nu(\omega) + \frac{1}{2} \right) = \frac{Q(\omega, r_{max})}{\left(2 Q''(r_{max})\right)^{1/2}}  =  \frac{\omega^2 - V(r_{max})}{\left(-2 V''(r_{max})\right)^{1/2}} \, ,
\ee
with $r_{max}$ being the maximum of $-Q(\omega, r) =   V(r) - \omega^2  $. When the potential and energy are real, one finds that $\left(\nu + 1/2\right)$ is imaginary. As a result
\be\label{T3}
(e^{i\pi\nu})^{*} =- e^{i\pi\nu}\,,\quad\, P^{*} = e^{-i\frac{\pi}{2}(\nu+1/2)}P^{-1} \, ,
\ee 
so that $T^{RW}$ indeed satisfies the transfer matrix conditions \eqref{eqn:TmatrixConditions}. From these, the Regge-Wheeler reflection and transmission coefficients can also we worked out
\be\label{T4}
|\mathcal{T}|^2 = \left(1+ e^{2 \pi i (\nu + 1/2)} \right)^{-1} \, , \quad |\mathcal{R}|^2 = \left(1+ e^{-2 \pi i (\nu + 1/2)} \right)^{-1} \, .
\ee
To third order in a WKB expansion, the following explicit expression for $P$ in terms of $\nu(\omega)$ can be found
  \be
 P = \left(\nu+\frac{1}{2}\right)^{(\nu+\frac{1}{2})}\exp\left[-\frac{1}{2}(\nu+\frac{1}{2})-\frac{1}{48} (\nu+\frac{1}{2})^{-1}\right]\,.
 \ee
Combining equations \eqref{NearhorBogo} and \eqref{ReggeWheelerconnection}, we find the advertised Bogoliubov relations between the asymptotic ingoing and outgoing modes 
\begin{align}\label{CompleteBogo}
 &\begin{pmatrix}
 S_{++}T^{RW}_{11}-T^{RW}_{21} & S_{+-}T^{RW}_{11}\\
 S_{+-}T^{RW}_{11} &  S_{++}T^{RW}_{11}-T^{RW}_{21}
 \end{pmatrix}\left( \begin{array}{c}
Z^{out \, (as)}_{\omega l m + } \\
Z^{out \, (as) }_{\omega l m - }  \end{array} \right) ~ = ~ \nn \\
&\qquad\qquad\qquad\qquad\quad = ~ \begin{pmatrix}
 T^{RW}_{22}-S_{++}T^{RW}_{12} & -S_{+-}T^{RW}_{12}\\
 -S_{+-}T^{RW}_{12} &   T^{RW}_{22}-S_{++}T^{RW}_{12}
 \end{pmatrix}
 \left(\begin{array}{c}
Z^{in  \, (as)}_{\omega l m + } \\
Z^{in \,  \, (as)}_{\omega l m - } \end{array} \right) \, . 
 \end{align}
This relation, as expected, is unitary. An important consequence of this relation is that the resulting matrix does not factorise 
\begin{equation}
\hat{S}_{total} = \hat{S}_{\mathcal{I}^- \rightarrow hor} \, \hat{S}_{horizon}  \, \hat{S}_{hor \rightarrow \mathcal{I}^+} 
\end{equation}
as was originally anticipated \cite{tHooft:1996rdg}. Instead, it is more complicated in form; modes that try to exit the near horizon region can backscatter against the Regge-Wheeler potential\footnote{A more detailed analysis of the behaviour of the effective potential in various regimes both at low and high frequency can be found in~\cite{Macedo:2013afa} and references within.}. Our result takes this backscattering into account, and we explore this further in Section \ref{ECHOES}.

A manifestly visible feature of the near horizon scattering matrix \eqref{eqn:SMatrix2b} is that the diagonal elements are exponentially suppressed in energy. This prevents the high frequency modes from scattering back to the same exterior of the Penrose diagram\footnote{After the antipodal identification of section. \ref{Antipodal}, this would mean that high frequency modes prefer to emerge from the other side of the black hole.}.  On the other hand low-energy modes are exponentially damped and cannot penetrate the Regge-Wheeler potential. These two effects conspire to ensure that highly energetic ingoing modes almost entirely pass from region $I$ to region $II$ while the low energy modes struggle to penetrate the Regge-Wheeler potential and almost entirely reflect back. This implies that the asymptotic scattering matrix \eqref{CompleteBogo} notwithstanding, as viewed from each exterior, the combined system predominantly behaves as if infalling boundary conditions were imposed near the horizon. This results in the expected quasi normal modes of the Schwarzschild black hole to leading order. This behaviour is corrected by exponentially small tails. As one traces their evolution back in time, Hawking modes in each exterior, are largely composed of modes that seem to emanate from the opposite exterior.

\section{Asymptotic S-matrix for one-sided black holes}\label{singleexteriorSmatrix}

\subsection{Antipodal identification}\label{Antipodal}

As discussed in the introduction, large single-sided black holes at late times long after collapse are well-approximated by an antipodally identified two-sided black hole \cite{Sanchez:1986qn, Hooft:2016itl, Hooft:2016cpw, Betzios:2016yaq, Strauss:2020rpb}. In global Kruskal coordinates, this identification is given by
\be
\Phi(U,V , \Omega) ~ = ~ \Phi^\dagger(-U, -V , \Omega^P) \, ,
\ee
and similarly, in the near horizon Eddington-Finkelstein coordinates \eqref{coord2}, we have
\be
\phi_+\left(u, v, \Omega\right) = \phi^\dagger_-\left(\bar{u}, \bar{v}, \Omega^P\right) \, . 
\ee
Here, $u, v$ refer to coordinates in region $I$ whereas the coordinates $\bar{u},\bar{v}$ describe coordinates in region $II$. The subscripts on the field refer to the regions occupied by the configurations. The identification is to be imposed on the near horizon modes of the extended Penrose diagram. We can then match these modes to those of a single Regge-Wheeler potential to cover the entire exterior region. Using the following relations for the spherical harmonics
\be
Y_{l \, m}^*(\Omega) = (-1)^m Y_{l \, -m}(\Omega)\, , \quad Y_{l\, m}(\Omega^P) = (-1)^l Y_{l \, m}(\Omega) \, ,
\ee
and imposing the antipodal identification, we obtain the following relations for the modes
\be
Z^{in (hor)}_{\omega, \, l, \, m, \, + }  \,  = \,  (-1)^{l - m} Z^{in (hor) \, *}_{-\omega, \, l, \, -m, \, - }   \, ,
\ee
with a similar relation for the outgoing modes near the horizon. Consequently, we find the following Bogoliubov relation for the near-horizon modes of a single-sided black hole 
\begin{equation}
\left( \begin{array}{c}
Z^{out (hor)}_{\omega, \, l, \, m} \\
Z^{out (hor) *}_{- \omega, \, l, \,  -m}  \end{array} \right) ~ = ~  \left( \begin{array}{cc}
S_{++}\left(\omega \right) & (-1)^{l+m}  S_{+-}\left(\omega \right) \\
(-1)^{l+m}  S_{-+}\left(\omega \right) & S_{--}\left(\omega \right) \end{array} \right) \left( \begin{array}{c}
Z^{in (hor)}_{\omega, \, l, \, m  } \\
Z^{in (hor) *}_{- \omega, \, l, \,  -m}  \end{array} \right) \, .
\end{equation}
Consistency of these equations with eqn. \eqref{eqn:SMatrix2b} gives
\be
Z^*_{- \omega, \, l, \,  -m}  = (-1)^{l+m} Z_{ \omega, \, l, \,  m} 
\ee
for both the in and outgoing modes. The resulting Bogoliubov relation simplifies further and is now given by the sum of the elements $S_{single}(\omega) = S_{++}(\omega) + S_{--}(\omega)$:
\be\label{singleS}
S_{one-sided}(\omega) = 2^{-i 4 G M \omega} e^{ -i 4 G M \omega \log \lambda_l} \frac{\Gamma \left(\frac{1}{4} - i 2 G M \omega \right)}{\Gamma \left(\frac{1}{4} + i 2 G M \omega \right)}  \, , \quad \lambda_l = \frac{8 \pi G}{R^2(l^2+l+2)} \, .
\ee
This is again a unitary relation; it is a pure phase. This one-sided result is due to the antipodal identification.

In \cite{Betzios:2020wcv} a refinement of the antipodal identification was proposed. It is one that additionally takes into account the near horizon backreaction commutation relations, briefly discussed in appendix~\ref{FirstQuant}. This results in a chaotic discrete spectrum with an extremely dense spacing of the order $\sim 10^{-80}$ (in Planck-units), even for a black hole of ten solar masses. Hence, we should emphasize that all the results in the following sections do not depend at all on such a possible microscopic discreteness, since it does not play any role for the low energy scattering experiments for which one can use continuous scattering states or wavepackets that cannot resolve such a fine-grained microstructure. More details can be found in appendix~\ref{FirstQuant}.

\subsection{The asymptotic S-matrix}\label{assmatrix}

\begin{figure}[t]
\begin{center}
\includegraphics[width=0.45\textwidth]{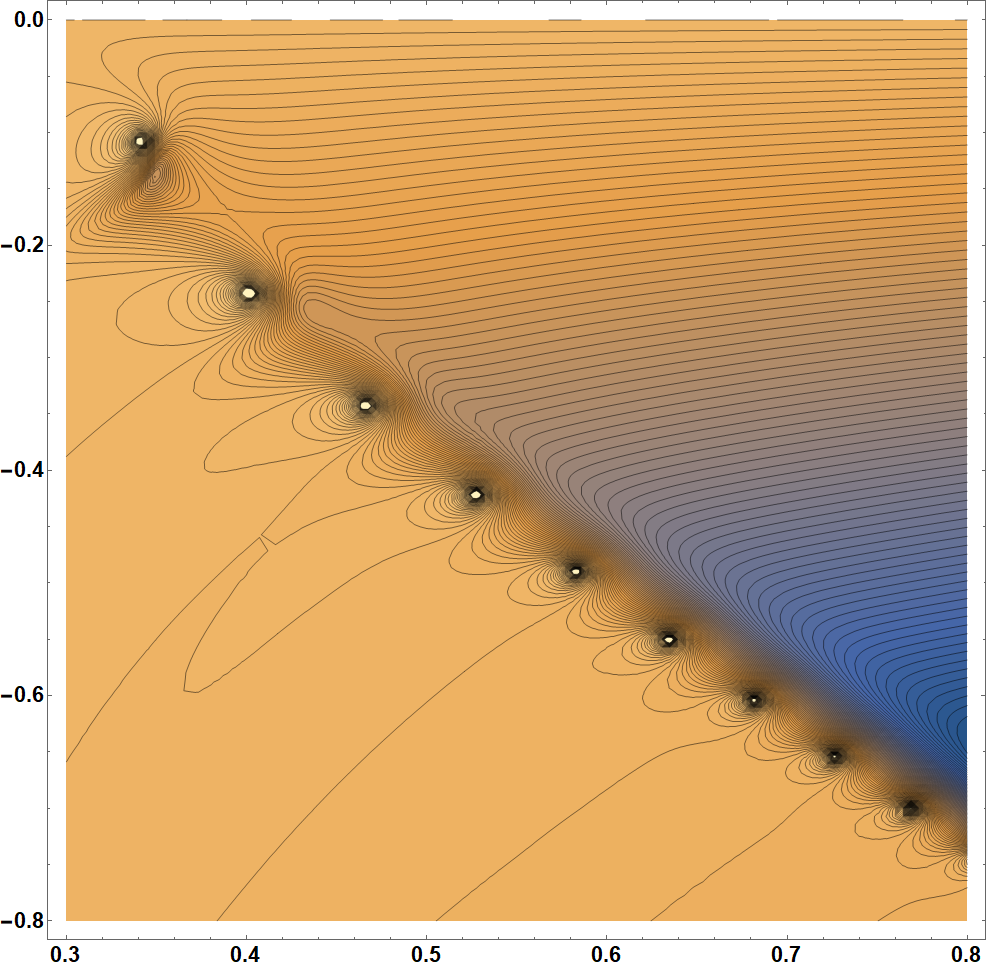} \hspace{0.1em} \includegraphics[width=0.05\textwidth]{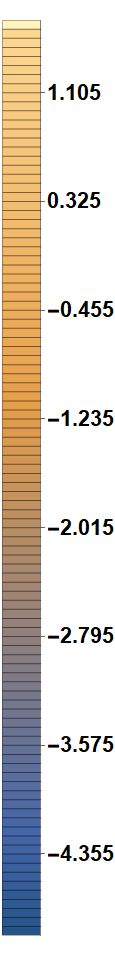} \hspace{0.1em} \includegraphics[width=0.45\textwidth]{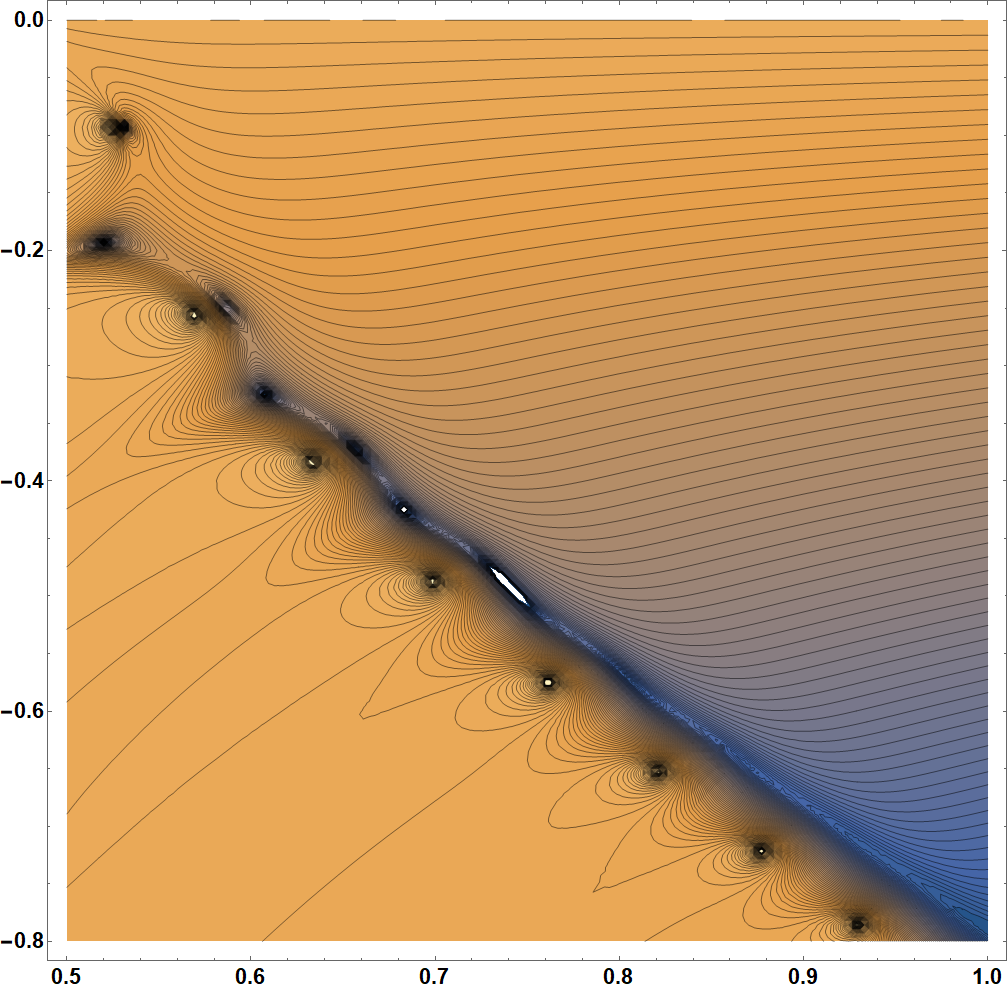} 
\end{center}
\caption{The two plots are contour plots of the logarithm of the absolute value of the S-matrix from eqn.~\eqref{singleS}. The plot is in the lower half complex $\omega$ plane (in Planck units) and the vertical axis parametrises $\Im \omega$ while the horizontal $\Re \omega$. In the left figure $l=1$, while in the right $l=2$. We find a series of resonances where the contours accumulate (dots). These resonances are what replaces the usual quasinormal modes of the black hole as described in the main text of subsection~\ref{assmatrix}. We also observe a deep trough where the S-matrix goes close to zero near the location of the poles, that becomes deeper and narrower for higher harmonics (blue region of the plot). These frequencies correspond to approximately trapped modes in the near horizon region.}
\label{fig:QNM1}
\end{figure}

After the antipodal identification, the asymptotic Bogoliubov relation for a single-sided black hole becomes much simpler. We use
\begin{equation}
\left( \begin{array}{c}
Z^{out \, (hor)}_{\omega l m  } \\
Z^{in \, (hor) }_{\omega l m  }  \end{array} \right) ~ = ~  \left( \begin{array}{cc}
T^{RW}_{11} & T^{RW}_{12} \\
T^{RW}_{21} & T^{RW}_{22} \end{array} \right) \left( \begin{array}{c}
Z^{out  \, (as)}_{\omega l m  } \\
Z^{in \,  \, (as)}_{\omega l m } \end{array} \right) \, .
\end{equation}
Together with $Z^{out \, (hor)}_{\omega l m } = S^{horizon} Z^{in \, (hor)}_{\omega l m }$ we find
\be\label{Ssingle}
Z^{out  \, (as)}_{\omega l m  } = \frac{S^{horizon}  T^{RW}_{22} - T^{RW}_{12}}{ T^{RW}_{11} - S^{horizon}  T^{RW}_{21}} Z^{in  \, (as)}_{\omega l m  }\, .
\ee
The poles of this Bogoliubov coefficient give the resonances of the scattering process; these are given by the condition
\be
T^{RW}_{11} =  S^{horizon}  T^{RW}_{21} \, .
\ee
The familiar quasinormal modes, on the other hand, are those for which $Z^{out \, (hor)}_{\omega l m  } = Z^{in \,  \, (as)}_{\omega l m } = 0 $. This is satisfied if $T^{RW}_{11} = 0$ yielding $\Gamma(- \nu) = \infty$, or that $\nu$ is an integer; see eqns. \eqref{T1}, \eqref{T2}. Therefore, the location of the unstable resonances approximately coincides with the location of the familiar quasinormal modes iff both $T^{RW}_{11} \approx 0$  and $S^{horizon} \approx 0$. A plot of the S-matrix resonances in the lower half complex $\omega$ plane for $l=1,2$ can be seen in fig.~\ref{fig:QNM1}.

It is of interest to compare these resonances to the familiar quasinormal modes of the Schwarzschild black hole. Our unitary S-matrix does not admit the usual notion of quasinormal modes (since it is not consistent to impose purely infalling boundary conditions on the horizon, due to the near horizon scattering relations). Nevertheless, we do find resonances in the complex lower-half plane, using the same asymptotic boundary conditions at infinity as those of the usual quasinormal modes (which is to have no infalling waves at infinity). The system can therefore dissipate only near asymptotic infinity, far away from the horizon. This means that such resonances continue to describe the equilibration of the black hole after it is brought into an excited state.

Furthermore, we observe the existence of deep troughs near these poles where the Bogoliubov coefficient \eqref{singleS} becomes approximately zero, or $S^{horizon}  T^{RW}_{22} \approx T^{RW}_{12}$ (their frequency also needs to be below the top of the Regge-Wheeler potential $\omega^2 < V_{max}$). They can be thought of as modes that are approximately trapped in the region between the horizon and the top of the Regge-Wheeler potential due to backscattering\footnote{Trapped modes have recently been discussed in the context of fuzzballs \cite{Bena:2019azk, Bena:2020yii}, but the difference with these works is that our background does contain a horizon. The mechanism of trapping is nevertheless analogous, due to the presence of an effective potential experienced by the modes, with a dip in the near horizon region.}.

The effect of backscattering has an important observable consequence: that of gravitational wave echoes to which we now turn. 

\section{Sources and echoes}\label{ECHOES}

Gravitational echoes during the emission of gravitational waves from the black hole have gained increasing interest \cite{Cardoso:2016rao, Cardoso:2016oxy, Cardoso:2019rvt, Abedi:2016hgu, Wang:2018gin}. In particular a simple picture of this effect is that of consecutive scattering back and forth of a wave in the near horizon region before it eventually manages to escape at infinity. Introduction of sources in the fluctuation equation excites the black hole and allows for a description of the subsequent emission of gravitation waves from the horizon. We shall closely follow the discussion in \cite{Mark:2017dnq,Cardoso:2019rvt}.

The Green's function for the Regge-Wheeler potential satisfies (in what follows, we drop the partial wave indices)
\be\label{RWG}
\left(\frac{\partial^2}{\partial r_*^2}  + \omega^2 - V(r_*) \right) G(r_*,r_*') = \delta(r_* - r_*') \, .
\ee
Using this, the field depending on a source $J$ is computed by
\be\label{RWG2}
\phi(r_*) = \int_{-\infty}^\infty d r_*' \, G(r_*,r_*') \, J(r_*') \, .
\ee
The two independent solutions to the homogeneous equation we shall need are the purely infalling one at the horizon (we call it $\phi_{inf}$)
\begin{align}\label{quasi2}
\phi^{(hor)}_{inf} ~ &\sim ~  e^{-i \omega r_*} \quad r_* \rightarrow - \infty \, , \nn \\
\phi^{(as)}_{inf}  ~ &\sim ~ Z_1^{out \, (as)} e^{i \omega r_*} + Z_1^{in \, (as)} e^{-i \omega r_*} \quad r_* \rightarrow + \infty\, .
\end{align}
and the purely outgoing one at asymptotic infinity (we call it $\phi_{up}$)
\begin{align}\label{out2}
\phi^{(hor)}_{up} ~ &\sim ~ Z_2^{out \, (hor)}  \, e^{i \omega r_*}  \, + \, Z_2^{in \, (hor)} \, e^{-i \omega r_*}  \quad r_* \rightarrow - \infty \, , \nn \\
\phi^{(as)}_{up} ~ &\sim ~ e^{i \omega r_*} \quad r_* \rightarrow + \infty \, .
\end{align}
In terms of these solutions, the construction of the Green's function of the quasi normal modes on the black hole background is given by
\be
G_{QNM}(r_*,r_*') = \frac{\phi_{inf}(r_*^{min}) \phi_{up}(r_*^{max})}{W_{BH}(\phi_{inf}, \phi_{up})} \, ,
\ee
where $W_{BH}$ is the Wronskian of the two solutions, and $min \, (max)$ refers to the minimum (maximum) of the coordinate $r_{*}$. This Green's function therefore is infalling at the horizon and outgoing at asymptotic infinity. The Wronskian is constant and can be computed to be $W_{BH}(\phi_{inf}, \phi_{up}) = 2 i \omega Z_2^{out \, (hor)}  $. We additionally have the following consistency relations $Z_2^{out \, (hor)} = Z_1^{in \, (as)}$ and  $Z_2^{in \, (hor)} = - Z_1^{out \, (as) \, *}$.

Instead of this quasi normal mode Green's function, we would like to define one that obeys the near horizon scattering relation
\be\label{nhboundary}
G^{horizon}_{S}(r_*,r_*') \sim   \, S^{horizon} \, e^{i \omega r_*}  \, +  \, e^{-i \omega r_*}  \quad r_* \rightarrow - \infty \, ,
\ee
so that it captures the near-horizon gravitational interactions. In order to do so, we may simply add to the quasi-normal mode solution another homogeneous solution of \eqref{RWG} times a function of $r_*'$, and demand that the resulting solution be consistent with the near horizon scattering boundary condition. The new Green's function can be written as follows
\be
G_{S}(r_*,r_*') = G_{QNM}(r_*,r_*') + \mathcal{K}(\omega, l) \frac{\phi_{up}(r_*^{min}) \phi_{up}(r_*^{max})}{W_{BH}(\phi_{inf}, \phi_{up})} \, .
\ee
In order to determine the coefficient $\mathcal{K}(\omega, l)$, we argue as follows. Let $r_* < r_*'$. We then obtain
\be
G_{S}(r_*,r_*') = \frac{\phi_{up}(r_*')}{W_{BH}} \left(\phi_{inf}(r_*)   +  \mathcal{K}(\omega, l) \phi_{up}(r_*) \right) \, ,
\ee 
so that close to the horizon $r_* \rightarrow - \infty$ we find
\be\label{dressedGreens}
G_{S}(r_*,r_*') = \frac{\phi_{up}(r_*')}{W_{BH}} \left(e^{-i \omega r_*} \left[ 1  + \mathcal{K}(\omega, l)  Z_2^{in \, (hor)}  \right]  + \mathcal{K}(\omega, l)  Z_2^{out \, (hor)}  \, e^{i \omega r_*}  \right) \, .
\ee 
Consistency with eqn. \eqref{nhboundary} gives
\be
\mathcal{K}(\omega, l) = \frac{S^{horizon}}{Z_2^{out \, (hor)}- Z_2^{in \, (hor)}  S^{horizon}} \, .
\ee
The final step is to rewrite the various coefficients in terms of the Regge-Wheeler transfer matrix (or scattering) elements of
eqns. \eqref{ReggeWheelerconnection}, \eqref{T1} and \eqref{RWreflectiontransmission}. 

We find $W_{BH} = 2 i \omega T^{RW}_{11} = 2 i \omega \mathcal{T}^{-1}$ and
\be\label{Kappa}
\mathcal{K}(\omega, l) = \frac{S^{horizon}}{T^{RW}_{11}- T^{RW}_{21}  S^{horizon}} \, = \, \frac{ \mathcal{T} S^{horizon}}{1 - \mathcal{R}  S^{horizon}}    \, .
\ee
The results are found to be consistent with those of section~\ref{singleexteriorSmatrix}. In particular the usual quasi-normal modes are given by the zeros of the Wronskian $W_{BH} = 2 i \omega T_{11}^{RW}$, whereas the expression for \eqref{Kappa} provides a different set of resonances. Possible echoes due to backscattering can be found by expanding the denominator of eqn. \eqref{Kappa} in a geometric series, to find
\begin{equation}\label{geometric}
 \mathcal{K}(\omega, l) ~ = ~  \mathcal{T} S^{horizon} \sum_{n=1}^\infty \left( \mathcal{R}  S^{horizon} \right)^{n-1} \, .
 \end{equation} 
 The interpretation is that after the main burst of the wave manages to pass the Regge-Wheeler barrier, an infinite number of consecutive waves are to be emitted as they scatter back and forth between the horizon and the Regge-Wheeler barrier. These are the ``echoes" of the main burst. Of course, if this interpretation is to hold, the echoes better be well separated in time. Differently said, the characteristic time width of the echo signal must be narrower than the time separation between consecutive echoes. Otherwise, we must use the complete Green's function given by eqn. \eqref{dressedGreens} and the signal acquires more complicated characteristics.

A rough estimate for the echo spacing $\Delta t_{echo}$ can be made as follows. The time delay is generically given by the Wigner-Smith time delay function of the S-matrix (see \cite{Betzios:2016yaq} for more details in the present context)
\be
\Delta t(\omega, l) = - i \hbar \frac{\partial}{\partial \omega} \log \det S^{horizon}(\omega, l)  \, . 
\ee
Now the consecutive echo spacing can be understood from the extra piece contained in each of the terms in the geometric expansion of the Green's function, that is
\be
\Delta t_{echo}(\omega, l) \, = \, - i  \frac{\partial}{\partial \omega} \log \det\left[ \mathcal{R} S_{horizon}(\omega, l) \right] \, = \Delta t_{scr}(\omega, l) + \Delta t_{back}(\omega, l)    \, .
\ee
We find that it contains two pieces. One is the time delay associated to the scattering event near the horizon; this is often called the \emph{scrambling time}~\cite{Sekino:2008he}, since it is a short timescale governing the fast ``effective scrambling of information" that falls towards the black hole. The other is associated to the reflection/backscattering off the Regge-Wheeler potential. Since this second piece is universal for all the models of echoes and has been analysed in detail in the literature, we shall focus on the specific \emph{scrambling time} of our model. It is frequency and harmonic dependent:
\bea
\frac{\Delta t_{scr}(\omega, l)}{\hbar} \, = \, 4 G M_{BH} \log \frac{R_{BH}^2 (l^2+l+1)}{16 \pi G} \, - \, 4 G M_{BH} \Re \, \Psi\left( \frac{1}{4} + i 2 G M_{BH} \omega \right)  \, . \nn \\
\eea
In this expression $\Psi(z)$ is the Di-Gamma function~\cite{DLMF} and we notice a term that is frequency independent (but harmonic dependent) and a term that is frequency dependent. Due to the large hierarchy $R_{BH}^2/G$, the first term gives the dominant contribution in the physical frequency regime. For very large $\omega$, we find
\begin{align}
\frac{\Delta t_{scr}(\omega, l)}{\hbar} \, &\sim \, 4 G M_{BH} \left( \log \frac{R_{BH}^2 (l^2+l+1)}{16 \pi G}  -  \log ( 4 G M_{BH} \omega ) \right) \nonumber \\
&\sim \, 4 G M_{BH} \log \frac{M_{BH}}{\omega } \, .
\end{align}
We observe that for $\omega < M_{BH}$ (which holds always since the black hole cannot radiate all its mass in a single burst and any source should be thought of as a small perturbation of the large black hole) this time is always positive. Similar forms for the scrambling time and the connection with echoes is described in~\cite{Cardoso:2016oxy,Abedi:2016hgu}. An advance of our approach is that there is no need for an arbitrary UV cutoff near the horizon as in models of exotic compact objects (ECO's), and that only physical IR parameters appear in the expression for the scrambling time. This is a hint of the universality of this expression, similar to that governing the chaos bound~\cite{Maldacena:2015waa} (but our expression depends additionally on the frequency $\omega$ of the mode). 

Another physical property of echoes is their damping factor. This is controlled by the modulus of the scattering factor. Since the near horizon scattering matrix $S_{horizon}$ is unitary by itself (for real frequencies), the damping is produced solely by the Regge-Wheeler potential. From the echo expansion eqn. \eqref{geometric}, we find that the damping factor of the amplitude between subsequent echoes is governed predominantly by $|Z|^{n+1}_{echo}/|Z|^n_{echo} \sim |\mathcal{R}|$, and therefore given by \eqref{T4} in the WKB approximation. For very large frequencies, $\mathcal{R} \rightarrow 0$ (the backscattering effect of the Regge-Wheeler potential becomes insignificant)  and so there are no echoes.

To summarise, we find that the late-time signal is dominated by the poles of the exact asymptotic S-matrix, whereas the prompt ringdown of the initial excitation/burst is governed by the dominant quasi-normal modes of the corresponding black hole spacetime. Of course, a generic source is not a monochromatic signal and has some width. It would be very interesting to study, in quantitative detail, the echoes of appropriate wavepackets such as those in appendix~\ref{Wavepackets}.

\section{Conclusions}\label{Conclusions}

In this paper, we have constructed a scattering matrix for a massless scalar field as observed by an asymptotic observer at spatial infinity. This relies on a combination of the near horizon scattering relation that captures strong gravitational interactions and the transfer matrix associated to the effective Regge-Wheeler potential. We find that the resulting S-matrix is unitary. In addition, the system admits resonances analogous to the familiar quasinormal modes of the Schwarzschild black hole, but for which the purely infalling boundary conditions at the horizon are replaced by the near-horizon scattering relation. Owing to the scattering back and forth of the modes between the horizon and the potential barrier, we find ensuing gravitational echoes that penetrate the effective potential barrier to reach the asymptotic observer. We also studied characteristic time scales associated to the scattering events and the echoes. There are several comments to be made about our results, in comparison to various aspects of literature which we turn to.

\paragraph{Gravitational Echoes} In the search of observational signatures from the near horizon scattering relation, we propose that it might be possible to find after-burst echoes due to backscattering effects in the region between the horizon and the effective potential barrier. The important novelty that our proposal brings is twofold. First, there is a direct physical motivation for the presence of echoes based on well-established physical laws (requiring only a combination of general relativity and quantum mechanics and not an analysis of microscopic UV physics), even though the classical geometric background is that of a usual black hole. Second, our proposal does not have any free adjustable parameters; everything is determined in terms of $\hbar$, the Planck mass $M_P$ and the black hole mass $M_{BH}$, in addition to the spectral profile of the excitation. This provides an excellent template for the search of gravitational echoes\footnote{At the time of writing of the present article, there is no observationally significant evidence of post-merger echoes~\cite{Tsang:2019zra, Abbott:2020jks}, but a model without free adjustable parameters like ours could perhaps help in increasing the confidence level of the statistics should echoes exist. On the other hand, it also means our proposal is falsifiable.}. 

That said, we have not studied quantum gravitational interactions in the most common physically realistic system of rotating Kerr black holes. Neither have we modelled the complicated setup of black hole mergers. Nevertheless, we expect that our model should be able to capture the universal characteristics of echoes (should they exist) for timescales well separated from the merger phase. This expectation is of course for the echoes that do not depend on specific rotational characteristics of the system. An interesting future direction is to generalise our construction using the results of \cite{BenTov:2017kyf}, in order to treat rotating black holes in the Kerr family.

\paragraph{Energy conditions and shockwaves} An intensive study of energy conditions applied to eternal black holes in the context of the AdS/CFT correspondence has revealed various interesting physical properties of black holes and shockwaves. The eternal black hole background corresponds to the thermofield double state of the dual field theory~\cite{Maldacena:2001kr}. In the case of asymptotically flat spacetime, one only has access to the thermofield double description in terms of bulk modes. The spacelike wormhole connecting the two exteriors is then a manifestation of the strong entaglement between the two QFTs in holography~\cite{Maldacena:2013xja}. From a bulk perspective this leads to the maximally entangled Hartle Hawking state with a smooth horizon. It is also clear that the Euclidean analytic continuation of the geometry is that of two disconnected cigar geometries that smoothly cap off in the interior and hence there is no Euclidean geometric connection. Sending positive null energy shockwaves $\langle T_{VV} \rangle > 0$ deforms the classical geometry in a way that does not allow the modes to cross the wormhole classically. In order for this to happen at the level of the classical geometry, one needs to couple the two field theories directly in a way such that shocks with negative null energy are created in the bulk $\langle T_{VV} \rangle < 0$, and the Hilbert space no longer factorises \cite{Gao, Bzowski:2018aiq, Bzowski:2020umc}. Some analogous Euclidean saddle point solutions are Euclidean wormholes connecting the two boundaries for which cross correlators are non trivial and can arise only by a UV soft coupling of the two QFT's~\cite{Betzios:2019rds}.

The present construction goes one step further in the sense that the notion of locality in the bulk fails by a small amount (for small causal diamonds) when both $\hbar$ and $G_N$ are non zero due to the uncertainty relations \eqref{B8} arising from gravitational scattering. In field theory it can be shown that the scattering arises due to an infinite resummation of gravitational soft interactions. In holography, this would correspond to an effect that is non-perturbative in $1/N$, that is not captured by the classical black hole geometric saddle. Such an effect may be captured in the Euclidean setting by the presence of introducing appropriate Euclidean wormholes in the path integral, so that the two cigar geometries no longer decouple. These wormholes would also not be classical saddle point solutions, since the energy conditions are not violated at this level. Furthermore, we do not find modes crossing between the two sides at the classical level. This is also corroborated by the field theoretic calculation of \cite{Gaddam:2020mwe, Gaddam:2020rxb}, that explicitly involves the resummation of infinite loop (ladder) diagrams. Let us also finally note that the single exterior counterparts of wormholes, after the antipodal identification, are ``vacuoles'' in space~\cite{Betzios:2017krj}.

\paragraph{Quantum chaos} An important question concerns the possibility of detecting chaotic behaviour in a field theoretic S-matrix~\cite{Polchinski:2015cea, Kitaev:2017awl, Rosenhaus:2020tmv}. Although a lot of models describing first quantised quantum mechanical systems do exhibit quantum chaotic properties in their S-matrix, in QFT it is much harder to establish such a behavior since it cannot be observed using perturbation theory.  What we find and argue instead is that such chaotic properties can be encoded in the effective Bogoliubov coefficients relating in/out modes in second quantisation. While the exponential sensitivity to the initial conditions is a feature of the near horizon S-matrix in Eddington-Finkelstein coordinates\textemdash a feature that is easy to demonstrate both in first and second quantisation~\cite{Polchinski:2015cea, Kitaev:2017awl, Rosenhaus:2020tmv}\textemdash it is much harder to understand chaos from a more direct spectral point of view (or from the point of view of the scattering matrix of the asymptotic observer).

What our study reveals is that it is nevertheless possible to define an auxiliary quantum mechanical system and relate it with the S-matrix. More precisely, the dynamics of this system provides the appropriate Bogoliubov relations for the second quantised modes of various fields. Signatures of chaotic behaviour can be encoded in these Bogoliubov coefficients. The previous argument about the need for non-perturbative effects is now satisfied, since the auxiliary system effectively takes into account non-perturbative resummation effects from the point of view of the second quantised theory as was recently shown in \cite{Gaddam:2020mwe, Gaddam:2020rxb}. To further corroborate this perspective, an important feature of the antipodal identification is that it brings forward the possibility of obtaining a discrete ``chaotic spectrum" for the black hole microstates~\cite{Betzios:2020wcv}. Taking a quotient with respect to similar discrete symmetries is known to produce chaotic spectra and provide relations to the Riemann zeta function~\cite{GutzwillerBook, Gutzwiller, Lax, Faddev, Betzios:2016lne, Savvidy:2018ffh}. It is clear that there is a large class of systems that could be analysed in a similar fashion. 

\paragraph{Microscopic models} An important distinction must be made between models for which chaos arises due to boundary conditions (like those of ``chaotic billiards'' or ``resonators") and models which exhibit chaotic properties due to explicit microscopic Hamiltonian dynamics (as is the case with the BFSS matrix model \cite{Banks:1996vh,Gur-Ari:2015rcq}). That said, there have been several proposals in the past arguing in favour of various possible microscopic models underlying the near-horizon gravitational scattering matrix \cite{Banks:2015mya, Betzios:2016yaq, Banks:2020zrt}, in terms of fundamental matrix/fermionic degrees of freedom. Nevertheless there has been no comprehensive analysis of such models, and neither has there been a matching with the low energy results for the scattering matrix\footnote{See \cite{Lam:2018pvp} though, for such an attempt in two dimensions.}. This is one of important outstanding problems that would directly address the UV properties of the S-matrix.

\section*{Acknowledgements}
It is a pleasure to acknowledge various helpful discussions with D. Anninos, A. Bzowski, A. Gnecchi, K. Papadodimas, M.M. Sheikh-Jabbari, G. 't Hooft, T. Tomaras, and N. Tsamis, on topics related to the black hole S-matrix. Special thanks go to E. Kiritsis for an important question that motivated us to formulate the combined picture presented in this article and to the anonymous referee for interesting questions that also helped us improve the clarity of the text. Additional acknowledgements go to Katerina Chatziioannou for correspondence related to gravitational waves and echoes.

P.B is supported by the Advanced ERC grant SM-GRAV, No 669288. N.G is supported by the Delta-Institute for Theoretical Physics (D-ITP) that is funded by the Dutch Ministry of Education, Culture and Science (OCW).

\begin{appendix}

\section{Units and conventions}\label{Conventions}

In the main text we use units for which $c=1$ keeping the factors of $\hbar, G$ explicit. To go to SI units one should replace $G M$ with $ G M/c^2$, for example the black hole radius is $R_s = 2 G M /c^2$. For a black hole of the solar mass this radius is approximately $3 \times 10^3$ meters, still much larger than the Planck scale (even though this is an extremely small black hole from an astrophysical point of view). One also finds the values  $\hbar = h/2 \pi$ with $h=6.626 \times 10^{-34} \, J\cdot s$ and the Planck length given by $\ell_P = \sqrt{\hbar G/c^3} = 1.616 \times 10^{-35} \, m$. These are both in SI units, with three digits accuracy. The entropy of the black hole is given by $S_{BH} = \pi R_s^2 / \ell_P^2 = A/4 \ell^2_P$ and the Hawking temperature by $T = \hbar c^3/8 \pi G M_{BH}$.
The last two useful formulae are the surface gravity 
\be
\kappa = \frac{1}{R_s} =  \frac{c^2}{4 G M} \, ,
\ee
and the energy of the asymptotic observer measured in units of surface gravity
\be
E_{as} = \frac{\omega}{\kappa} = \frac{4 GM}{c^2 } \omega \, .
\ee

\subsection{Metric conventions}

From now on we work in $c=1$ units to simplify our formulae.

\begin{figure}[t]
\centering
\includegraphics[width=0.7\textwidth]{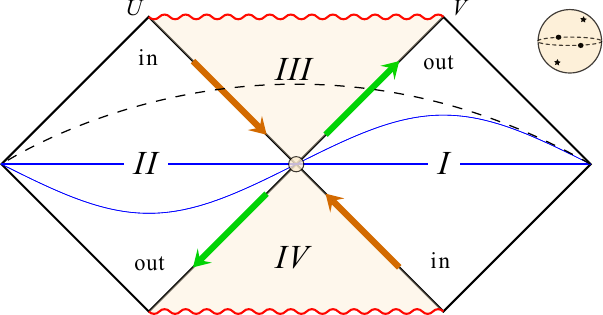}
\caption{The Penrose diagram and the antipodal map as originally proposed in~\cite{Sanchez:1986qn,Hooft:2016itl,Hooft:2016cpw,Betzios:2016yaq}. The original $t=0$ Cauchy slide that defines the thermofield double is drawn in blue. We depict two possible choices of evolving the slices and quantisation. One is using the Killing vector fields of the geometry with which the Cauchy slices change as shown in the thin blue curve. In contrast, the standard global time evolution involves the slicing drawn with dashed lines. After the antipodal identification, we can remove regions $II$ and $IV$. The bifurcation $S^2$ is then antipodally identified as shown in the top right corner. The first type of slices then contain always the complete information for the state of the black hole and pass from the antipodally identified bifurcation two sphere. In the second case instead, a part of the slice is always outside of the horizon.}
\label{fig:Penroseflat}
\end{figure}

We take the Schwarzschild metric to be of the form
\be
ds^2 = - \frac{3 2 G^3 M^3 e^{- r/2GM}}{r}  dU \, d V  \, + \, r^2 d \Omega^2 \, , \quad U V = \left( 1 - \frac{r}{2 G M} \right) e^{r/2GM}
\ee
The extended Penrose diagram is completely covered with the coordinates $U, V, \Omega$ but we can also split them as follows
\begin{align}
U ~ &= ~ - \frac{1}{4 G M} e^{-u/4 G M} \, , \quad U ~ = ~ + \frac{1}{4 G M} e^{-\bar{u}/4 G M} \, , \nonumber \\
V ~ &= ~ + \frac{1}{4 G M}  e^{+v/4 G M} \, , \quad V ~ = ~ - \frac{1}{4 G M} e^{+\bar{v}/4 G M} \, , \label{coord2}
\end{align}
in terms of the $u = t - r^*$, $v= t + r^*$ outgoing/ingoing Eddington-Finkelstein coordinates in region $I$ (for which $U<0,\, V>0$) and similarly using the coordinates $\bar{u},\, \bar{v}$ in region $II$ (for which $U>0,\, V<0$). Notice that we define the asymptotic regions of the fully extended Penrose diagram using the Killing vector fields of the geometry that have an opposite orientation at the two sides of the diagram. For example $\mathcal{I}^+$ is the union of the upper right and lower left null lines of the diagram~\ref{fig:Penroseflat}. These are the conventions used in~\cite{Hooft:2015jea,Betzios:2016yaq,Hooft:2016itl,Betzios:2020wcv} when defining the scattering problem on the extended Kruskal manifold.

The near horizon geometry is that of Rindler spacetime. To see this, one can expand the metric close to $r = 2 GM$ using
\be
r = 2 G M + \frac{x^2}{8 G M}
\ee
to find Rindler space
\be
ds^2_{N.H.} = - \kappa^2 x^2 dt^2 + dx^2 + (2 G M)^2 d \Omega^2
\ee
or in Kruskal-like coordinates
\be
ds^2_{N.H.} = -  dU \, d V  \, + \, (2 G M)^2 d \Omega^2 \, .
\ee
This is the near horizon geometry on which the shockwave backreaction is the leading physical effect on the S-matrix.
The lines $U=V=0$ generate the bifurcate Killing horizon of $k = \partial/\partial t$. In $(U,V)$ coordinates it takes the form
\be
k = \kappa \left(V \frac{\partial}{\partial V} - U \frac{\partial}{\partial U} \right) \, .
\ee
The orbits have opposite orientation in the regions $I, II$ of the Penrose diagram as depicted in fig.\ref{fig:Penroseflat}.

\section{The various vacua}\label{Vacua}

In this appendix, we provide some details on the three most common states/vacua employed in the study of the Schwarzschild black hole together with some comments related to the present work:

\begin{itemize}

\item The Hartle-Hawking-Israel state $| H \rangle$~\cite{Israel:1976ur}, which is smooth at both the past and future horizon. To achieve this incoming modes are positive frequency with respect to $V$ which is the affine parameter at the future horizon and outgoing modes are of positive frequency with respect to $U$ which is the affine parameter on the past horizon. Such a vacuum can be defined in Minkowski space and the eternal Kruskal manifold but not in a collapsing scenario. It has both the past and future horizons smooth and is a highly tuned equilibrium state. Nevertheless if the black hole is large, after some time the Hartle-Hawking vacuum starts resembling the Unruh vacuum, especially from the point of view of the infalling observer that perceives a smooth horizon.

\item The Boulware $| B \rangle$ vacuum~\cite{Boulware:1974dm} which is defined by requiring the normal modes to be of positive frequency with respect to the Killing vector $\partial/\partial t$ with respect to which the exterior region is static. It is hence a natural choice of vacuum state from the point of view of the external observer (no particles at asymptotics), for which the scattering process takes place. In this vacuum the stress energy tensor in both the future and past horizons is singular. It indicates that one should properly take into account the backreaction of the Hawking particles as one traces back their evolution from $\mathcal{I}^+$ to the horizon where they suffer a large blueshift effect.

\item The Unruh vacuum~\cite{Unruh:1976db} $| U \rangle$, which should be thought of as an intermediate situation of the previous two. It is defined by requiring the modes coming from $\mathcal{I}^-$ to be of positive frequency with respect to $\partial/\partial t$, while those that emanate from the past horizon $\mathcal{H}^-$ are of positive frequency with respect to $U$ which is the affine parameter on the past horizon. It is singular on the past horizon. It also indicates a large backreaction effect once one traces the modes back to the past horizon from which they emanate from. This has been argued in the past to be the most realistic vacuum for the late time scattering on a black hole background formed by collapse~\cite{Wald:1999vt}.

\item The antipodally identified\footnote{Or the corresponding state in the CPT gauged theory \cite{Betzios:2020wcv}.} state~\cite{Hooft:2015jea,Betzios:2016yaq,Hooft:2016itl,Hooft:2019nmf,Betzios:2020wcv}. It can also be understood as a form of formal boundary state~\cite{Bzowski} arising after the antipodal identification of the Hartle-Hawking-Israel state. One can also naturally regulate this identification at a Planckian distance from the horizon~\cite{Betzios:2020wcv}, which was argued to be a consequence of making the CPT symmetry a local symmetry. It is still an open question whether the typical state naturally formed in the process of gravitational collapse resembles this state, but this would require Planckian scale physics to determine. In this article we shall take the perspective that this state is a good representative of the state formed by collapse for timescales that are later than the formation time and earlier than the evaporation time so that we can use the hierarchically large parameter $M_{P}/M_{BH}$ to perform our computations.

\end{itemize}

\section{Properties of Bogoliubov transformations}\label{Bogoliubov}

Consider an S-matrix relation between asymptotic states in a second quantised theory
\be
| \Psi \rangle_{out} \,  = \, \hat{S} \, |\Psi \rangle_{in} \, ,
\ee
and suppose that we can expand the quantum field in terms of field modes and creation/annihilation operators as
\bea
\hat{\phi} \, = \, \sum_{J} \hat{a}^{in}_{J} \phi_J^{in} \, + \,  \hat{a}^{\dagger \, in}_{J} \phi_J^{* \, in} \, = \, \sum_{J} \hat{a}^{out}_{J} \phi_J^{out} \, + \, \hat{a}^{\dagger \, out}_{J} \phi_J^{* \, out} \, , \nn \\
\hat{a}^{in}_{J} | 0 ; in \rangle \, = \, 0 \, , \quad \hat{a}^{out}_{J} | 0 ; out \rangle \, = \, 0 \, .
\eea
Generically the asymptotic in/out field modes will be related via a Bogoliubov transformation\footnote{In the most general case, this is not always in terms of a square matrix, and merely expresses a canonical relation between different phase space variables.}. Assuming that the in/out modes are properly normalised the Bogoliubov coefficients are given in terms of the Klein Gordon inner product
\begin{align}
A_{J J'} \, &\equiv \, \langle \phi_{J'}^{in} , \, \phi_J^{out}  \rangle \, , \nonumber \\
B_{J J'} \, &\equiv \, - \langle \phi_{J'}^{in \, *} , \, \phi_J^{out}  \rangle \, , \nonumber \\ 
\langle \phi^1 , \, \phi^2  \rangle \, &= \, i \int_\Sigma d \sigma^\mu  \left( \phi^{1 \,*} \phi^2_{; \mu} - \phi^{2} \phi^{1 \, *}_{; \mu}  \right)  \, .
\end{align}
In this formula $\Sigma$ denotes the hypersurface on which the product is evaluated and $\sigma^\mu$ is a future directed vector normal to the hypersurface.

In terms of the S-matrix, this results to the following relation between creation and annihilation operators and modes
\bea
 \hat{S} \, \hat{a}^{in}  \, \hat{S}^\dagger \, \equiv \,  \hat{a}^{out}  \quad , \quad   \phi_J^{ out} =  \sum_{J'} \,   A_{J J'}  \,  \phi_{J'}^{in} \, + \,  B_{J J'} \,  \phi_{J'}^{* \, in}   \, ,  \nn \\
\begin{pmatrix}
\phi^{ out} \\ \phi^{* \, out}
\end{pmatrix} \, =  \, C \begin{pmatrix}
\phi^{ in} \\ \phi^{* \, in}
\end{pmatrix} \, , \quad C = \begin{pmatrix}
A & B \\ B^* & A^*
\end{pmatrix} \, , \quad C^{-1} = \begin{pmatrix}
A^\dagger & -B^T \\ -B^\dagger & A^T
\end{pmatrix} \, ,
\eea
where in the second line we used an implicit matrix notation (this includes any continuous labels to be integrated over.  $C^{-1}$ is used to find the inverse Bogoliubov transformation. The canonical property of the transformation leads to the following relations for the submatrices 
\bea
A A^\dagger - B B^\dagger = 1 \, , \quad A^\dagger A - B^T B^* = 1 \, , \nn \\
(A^{-1} B)^T = A^{-1} B \, , \quad (B^* A^{-1})^T = B^* A^{-1} \, .
\eea

The S-matrix can be written as (the expression is to be normal-ordered)
\bea
\hat{S} &=& \det \left( A^\dagger A \right)^{- 1/4} \normord \exp \left(\half  a_{I}  M^*_{I J} a_{J} +  a_{I}  Q^*_{I J} a^\dagger_{J} + \half  a^\dagger_{I}  P^*_{I J} a^\dagger_{J}  \right) \normord \, , \nn \\
\hat{S}^\dagger &=& \det \left( A^\dagger A \right)^{- 1/4}  \normord \exp \left(\half  a_{I}  P_{I J} a_{J} +  a_{I}  Q_{I J} a^\dagger_{J} + \half  a^\dagger_{I}  M_{I J} a^\dagger_{J}  \right) \normord
\eea 
where we adopt a summation convention, supress any continuous labels, and where we employ the (symmetric for bosonic fields) matrices
\be
M_{I J} = -  \left[B^* A^{-1}  \right]_{I J} \, , \quad P_{I J} =   \left[ A^{-1} B  \right]_{I J}   \, , \quad  Q_{I J} = \left[ A^{-1} - \mathbb{I}  \right]_{I J} \, .
\ee
The matrices $P,M$ and $Q$ govern the creation/annihilation and scattering of particles.

One can also express the out-vacuum in terms of the in-vacuum as
\be
|0 ; out \rangle \, = \, \hat{S} \, |0 ; in \rangle \,   = \, \det \left( A^\dagger A \right)^{- 1/4} \, \exp \left( \half  a^\dagger_{I}  P^*_{I J} a^\dagger_{J}  \right) \, |0 ; in \rangle  \, .
\ee

\section{Wavepackets}\label{Wavepackets}

In order to regulate any possible divergences that arise when considering computations using the delta function normalised partial wave modes and to bring the results into contact with experiments that inevitably have a finite resolution, it is convenient to use instead the scattering of localised wavepackets. Hawking introduced the following orthonormal basis
\be
\phi_{j n l m} = \frac{1}{\epsilon^{1/2}} \int_{j \epsilon}^{(j+1)\epsilon} d \omega \, e^{2 \pi i n \omega /\epsilon} \, \phi_{\omega l m}  \, , 
\ee
where $\epsilon = \delta \kappa$ with $\kappa$ the surface gravity and $0 < \delta \ll 1$. In these packets the frequencies are in the window $[j \epsilon, \, (j+1) \epsilon]$, with a width of $\Delta u \sim 2 \pi/\epsilon$ and centered around $u = 2 \pi n /\epsilon$ for outgoing packets at $\mathcal{I}^+$. Similarly expansions can be performed in the rest of the regions. In order not to clutter the notation we will use a colective index $J \equiv \lbrace j n l m \pm \rbrace$, where we introduce again the two signs in order to denote the two black hole regions in case we wish to describe the extended Kruskal manifold. Such packets are orthonormal with respect to the Klein-Gordon inner product as
\be
\langle \phi_J , \, \phi_{J'} \rangle = - \langle \phi^*_J , \, \phi^*_{J'} \rangle =  \delta_{J J'} \, , \quad \delta_{J J'} = \delta_{n n'} \delta_{j j'} \delta_{l l'}  \delta_{m m'} \, .
\ee
For late time wavepackets $n \geq N \gg 1$, so that they are peaked at large retarded time. We can either introduce mixed continuous/discrete Bogoliubov coefficients given by
\be
A_{j n \omega'} = \frac{1}{\epsilon^{1/2}} \int_{j \epsilon}^{(j+1)\epsilon}  d \omega \, e^{2 \pi i n \omega /\epsilon} \, A_{\omega \omega'}  \, , \quad B_{j n \omega'} = \frac{1}{\epsilon^{1/2}} \int_{j \epsilon}^{(j+1)\epsilon} d \omega \, e^{2 \pi i n \omega' /\epsilon} \, B_{\omega \omega'} \, ,
\ee
or pass to a completely discrete basis $A_{j n j' n'}, \, B_{j n j' n'} $ employing a double integral transform on both frequency indices
\bea
A_{j n j' n'} = \langle \phi^{in}_{J'} , \, \phi^{out}_{J}   \rangle =  \frac{1}{\epsilon}  \int_{j \epsilon}^{(j+1)\epsilon}  d \omega  \int_{j' \epsilon}^{(j'+1)\epsilon}  d \omega' \, e^{2 \pi i ( n \omega \, - \, n' \omega')/\epsilon } \,  A_{\omega \omega'} \, , \nn \\
B_{j n j' n'} = - \langle \phi^{in \, *}_{J'} , \, \phi^{out}_{J}   \rangle =  \frac{1}{\epsilon}  \int_{j \epsilon}^{(j+1)\epsilon}  d \omega  \int_{j' \epsilon}^{(j'+1)\epsilon}  d \omega' \, e^{2 \pi i ( n \omega \, + \, n' \omega')/\epsilon } \,  B_{\omega \omega'} \,.
\eea
If the Bogoliubov transformation is diagonal in the frequency basis i.e. $A_{\omega \omega'} = A_\omega \delta(\omega - \omega')$ the expression simplifies into
\begin{align}
A_{j n j n'} ~ &= ~ \frac{1}{\epsilon}  \int_{j \epsilon}^{(j+1)\epsilon}  d \omega  \, e^{2 \pi i \omega ( n  \, - \, n')/\epsilon } \,  A_{\omega} \, , \nonumber \\ B_{j n j n'} ~ &= ~ \frac{1}{\epsilon}  \int_{j \epsilon}^{(j+1)\epsilon}  d \omega  \, e^{2 \pi i \omega ( n  \, + \, n')/\epsilon } \,  B_{\omega} \, .
\end{align}

In particular since we wish to focus on late time scattering in which the details of the collapse do not matter, we can study the limit $j \gg 1 $, so that we can use Stirling and a saddle point approximation for the integral. For the near horizon shockwave relation eqn. \eqref{singleS} we find
\begin{align}
 A_{j n j n'} ~ &\simeq ~ \frac{1}{\epsilon}   \int_{j \epsilon}^{(j+1)\epsilon}  d \omega  \, e^{2 \pi i \omega ( n  \, - \, n')/\epsilon } e^{- i \omega 4 G M (\log 2 \lambda \, - 1)} e^{- i 4 G M \omega \log 4 G M \omega}  \nonumber \\
 &\simeq ~ \frac{1}{\epsilon}   \frac{\sin  \epsilon z /2}{z} \quad \text{with} \, \, z = \frac{2 \pi (n - n')}{\epsilon} - 4 G M (\log 2 \lambda \, - 1) - 4 GM \log 4 G M j \epsilon \, .
\end{align}
This has the form of a shifted \emph{sine-kernel}, since the maximum does not appear exactly on the diagonal $n=n'$.
In the two-sided case we can perform a similar analysis, the main difference being that there is a large exponential suppression for the packet to emerge on the same side of the Penrose diagram.

In general useful orthonormal bases of wavepackets are the so called wavelet orthonormal bases. It might be interesting to study such a basis, since the eigenstates exhibit special properties under Dilations.

\section{The first quantised description}\label{FirstQuant}

In this appendix, we briefly review the first quantised description used in~\cite{Hooft:2015jea, Betzios:2016yaq, Hooft:2016itl, Betzios:2020wcv}.

In \cite{Dray:1984ha}, Dray and 't Hooft proposed a simple method to properly take into account gravitational backreaction effects that is based on the study of the scattering of Aichelburg-Sexl gravitational shockwaves in the near horizon region of the black hole. This treats ingoing and outgoing particles as properly boosted small black holes, the gravitational field of which is mutually affecting their evolution as they cross each other. The effect is encapsulated in a lightcone shift for the trajectories of the scattered particles. On the other hand this shockwave field can be kept small enough compared to the total field of the large macroscopic black hole, so that one can consider a quasi-adiabatic approximation, in which such particles mutually scatter on a background of an approximately fixed total radius $R_{BH}$ and mass $M_{BH}$. The adiabatic approximation in particular is based on the fact that the rate of change of the parameters describing a large black hole is parametrically smaller compared to the faster timescales involved in the scattering process of individual modes; the later is sometimes called the \emph{scrambling} time as we describe in the main text. Using these assumptions, the backreaction calculation of 't Hooft gave rise to the following formal expression for the shockwave S-matrix (in the following expressions the coordinates and momenta are dimensionless variables)
\be\label{B6}
\langle P_{out} ; \Omega' | \hat{S}_{shock} | P_{in} ; \Omega \rangle = \exp \left( - i \frac{8 \pi G}{R^2} \int d^2 \Omega \int d^2 \Omega' P_{out}(\Omega) f(\Omega, \Omega') P_{in}(\Omega') \right) \, ,
\ee
where $P_{in/out}(\Omega)$ are lightcone momentum distributions of ingoing and outgoing particles on the $S^2$.
An equivalent description of this scattering process is through the relations
\begin{align}\label{B7}
V^{in}(\Omega) ~ &= ~ - \frac{8 \pi G}{R^2} \int d \Omega' f(\Omega, \Omega')P_{out}(\Omega') \, , \nn \\ 
U^{out}(\Omega) ~ &= ~ \frac{8 \pi G}{R^2}  \int d \Omega' f(\Omega, \Omega')P_{in}(\Omega') \, , \nn \\ 
(\nabla_{\Omega} - 1) f(\Omega, \Omega') ~ &= ~ - \delta(\Omega-\Omega') \, .
\end{align}
It was subsequently understood~\cite{Hooft:2015jea} that since the kernel $f(\Omega, \Omega')$ obeys a two-dimensional Laplace like equation, a spherical harmonics expansion can simplify the analysis of the S-matrix, and gives rise to the scattering amplitude of eqn.~\ref{eqn:amplitude} in the main text, as well as to the following algebra upon quantising the position and momentum distributions
\be\label{B8}
[\hat{V}^{in}_{l m}, \, \hat{U}^{out}_{l' m'} ] = i \hbar \lambda_l \delta_{l l'} \delta_{m m'} \, , \qquad \lambda = \frac{8 \pi G }{ R^2 (l^2 + l + 1)} \, .
\ee
This algebra is quite intriguing, since it elucidates the quantum gravitational nature of the effect (It trivialises as $G/R^2 \rightarrow 0$ and reduces to poisson bracket relations at the classical level when $\hbar \rightarrow 0$). Another important aspect of this algebra is that it indicates an inherent uncertainty in determining the exact lightcone position of propagating modes since~ $\Delta V \,  \Delta U \, \geq \, \hbar \, \lambda$, there exist minimal size causal diamonds for each harmonic centered around $U=V=0$. The spacetime then acquires some form of non-commutativity, slightly different though from the usual definitions of non-commutative geometry since the non-commutativity parameter depends on the presence of the transverse space through the harmonic number $\ell$. One would expect this non-commutative structure to be an important feature of Planckian sized black holes, for which $\hbar G/R^2 \sim O(1)$, but it is also conceivable that the algebra could be modified in this regime due to additional $1/M_P$ effects, that could arise from a more microscopic, possibly string theoretic description. Furthermore, it is also tempting to regard the two-dimensional Euclidean horizon as some form of string worldsheet or brane where $P_{in/out}(\Omega)$ play the role of vertex operators that describe the ingoing/outgoing particles. 

Regardless of whether such a string theory picture is ultimately correct, it is straightforward to describe the state of the black hole in a first quantised fashion in terms of wavefuctions that depend on the positions or momenta of the partial waves i.e. $\langle V ; l m | \Psi^{in} \rangle \equiv \Psi^{in}_{l m}(V) \,$,$\, \langle U ; {lm} | \Psi^{out} \rangle \equiv \Psi^{out}_{l m}(U)$. The algebra and the S-matrix relation \eqref{B6} then lead to a corresponding Fourier transform relation between such in/out lightcone wavefunctions
\be\label{B9}
\Psi^{out}_{l m}(U) = \int_{-\infty}^\infty  \frac{d V}{\sqrt{2 \pi \hbar \lambda_l}} e^{- \frac{i}{\hbar \lambda_l} U V}  \Psi^{in}_{l m}(V)
\ee
that inevitably leads to an explicit unitary near horizon S-matrix as discussed in more detail in~\cite{Hooft:2015jea,Betzios:2016yaq,Hooft:2016itl,Betzios:2020wcv}. 

\begin{figure}[t]
\centering
\includegraphics[width=0.5\textwidth]{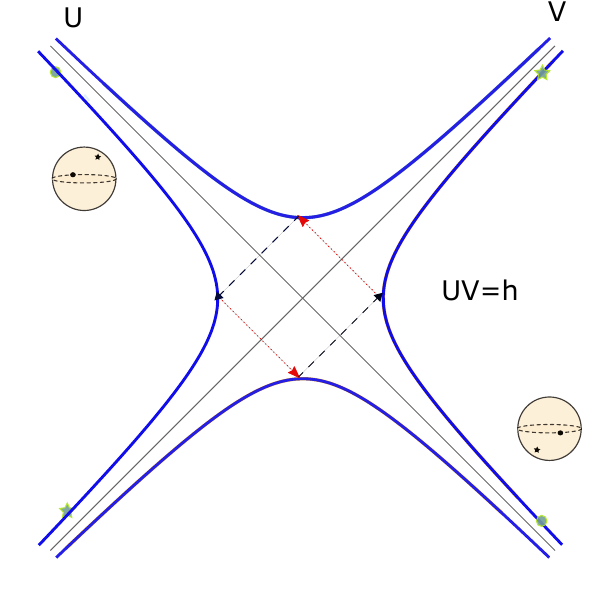}
\caption{The near horizon local CPT identification proposed in~\cite{Betzios:2020wcv}. We use the double sided Penrose diagram only to describe the near horizon region (pink region of fig. \ref{fig:Regions}) and we perform the identification up to the hyperbolae that are at a Planckian distance from the horizon, as in~\cite{Betzios:2020wcv}.}
\label{fig:antipodal}
\end{figure}

In \cite{Betzios:2020wcv} a refinement of the antipodal identification was proposed. It is one that additionally takes into account the near horizon backreaction commutation relations eqn. \eqref{B8}.
In short, it is not possible to apply the identification all the way up to $U=V=0$, since there is a backreaction algebra and phase space structure that needs to be properly taken into account. Once this is done, the resulting application of the identification can only hold up to the Planckian hyperbolae $U V = h$, as shown in fig.~\ref{fig:antipodal}. The boundary condition precisely on these Planckian hyperbolae results in a quantisation condition for the allowed black hole spectrum, that is given by the Riemann zeros (shifted by a constant term that depends on the harmonic~\footnote{The energetic degeneracy of even odd modes (with respect to Kruskal coordinates) of the near horizon region is now lifted~\cite{Betzios:2020wcv}.}). In Planck units it is found to behave as
\be\label{densespectrum}
\omega_l = M_{BH} \left( \frac{M_P^2}{4  M^2_{BH}}  E + \frac{(l^2 + l + 1)}{8 \pi } \right) \, , \quad \zeta \left(\frac{1}{2} \pm i E \right) = 0 \, .
\ee
We observe that the allowed spectrum is extremely dense since the Riemann zeroes are multiplied by a very small number (for example $(M_P/M_{BH})^2 \sim 10^{-80}$ even for small black holes of approximately 10 solar masses). Even though this suggests the possibility of a fundamentally discrete spectrum for the black hole microstates, due to the extremely dense spectrum, we can treat it as approximately continuous to describe the scattering process for any low energy asymptotic observer. This is the reason why none of the results in this article does depend on such a possible microscopic discreteness that is unobservable for ordinary probes.

In this work we emphasised instead on a second quantised field theoretic interpretation of the shockwave scattering relations eqns.~\eqref{B6} and~\eqref{B7}. From this perspective the wavefunctions are now simply mode functions of a spinless/massless scalar field. The fourier transform relation provides a \emph{near horizon mode relation} that can be used to extract the relevant Bogoliubov coefficients. This mode relation modifies Hawking's WKB approximation encapsulated by eqn.~\ref{WKBrel}, and gives rise to a unitary relation. This relation is precisely what replaces Hawking's WKB geometric optics approximation. It has the appealing feature that the UV problem of the infinite blueshift is tamed through the lightcone fourier transform relation; highly energetic and localised ingoing modes are scattered and replaced by soft delocalised outgoing modes and vice versa. 

Since the shockwave equations originate from an analysis of Einstein's equations in the presence of shockwave sources, they are inherently on-shell field theoretic relations bearing a direct effect on the propagating modes in the near horizon region. The second quantised generalisation which we used in this article also include off-shell fluctuations \cite{Gaddam:2020mwe, Gaddam:2020rxb}.

To summarise, eqns. \eqref{B7} and \eqref{B9} are what replaces eqns. \eqref{WKBrel}, but now these are merely \emph{near horizon} and not asymptotic relations for the modes. In sections~\ref{twoexteriors} and~\ref{singleexteriorSmatrix} of the main text, we show how they can be appropriately extended to \emph{asymptotic} relations for the modes of a scalar field.

\end{appendix}

%\addcontentsline{toc}{section}{Bibliography}

%\bibliography{bibuscodif} % texto.bib es el fichero donde está salvada la bibliografía.
%\bibliographystyle{unsrt} % estilo de la bibliografía.
\end{document}